\documentclass[showpacs,pra,aps,nopacs,onecolumn,twoside,superscriptaddress]{revtex4}

\usepackage{amsmath,amsfonts,amssymb,caption,color,epsfig,graphics,graphicx,hyperref,latexsym,mathrsfs,revsymb,theorem,url,verbatim,epstopdf}

\hypersetup{colorlinks,linkcolor={blue},citecolor={red},urlcolor={blue}}
\usepackage{subfigure}
\pdfoutput=1

\newtheorem{definition}{Definition}

\newtheorem{theorem}[definition]{Theorem}

\def\squareforqed{\hbox{\rlap{$\sqcap$}$\sqcup$}}
\def\qed{\ifmmode\squareforqed\else{\unskip\nobreak\hfil
\penalty50\hskip1em\null\nobreak\hfil\squareforqed
\parfillskip=0pt\finalhyphendemerits=0\endgraf}\fi}
\def\endenv{\ifmmode\;\else{\unskip\nobreak\hfil
\penalty50\hskip1em\null\nobreak\hfil\;
\parfillskip=0pt\finalhyphendemerits=0\endgraf}\fi}

\def\Dbar{\leavevmode\lower.6ex\hbox to 0pt
{\hskip-.23ex\accent"16\hss}D}
\makeatletter
\def\url@leostyle{%
  \@ifundefined{selectfont}{\def\UrlFont{\sf}}{\def\UrlFont{\small\ttfamily}}}
\makeatother
\newcommand{\bra}[1]{\langle#1|}
\newcommand{\ket}[1]{|#1\rangle}

\newcommand{\braket}[2]{\langle#1|#2\rangle}

\def\Dbar{\leavevmode\lower.6ex\hbox to 0pt
{\hskip-.23ex\accent"16\hss}D}

\begin{document}\large

\title{Multiqubit entanglement due to quantum gravity}

\date{\today}

\author{Shaomin Liu}
\affiliation{School of Mathematics, Physics and Finance, Anhui Polytechnic University, Wuhu 241000, China}

\author{Lin Chen}\email[]{linchen@buaa.edu.cn (corresponding author)}
\affiliation{LMIB(Beihang University), Ministry of Education, and School of Mathematical Sciences, Beihang University, Beijing 100191, China}
\affiliation{International Research Institute for Multidisciplinary Science, Beihang University, Beijing 100191, China}
\author{Mengfan Liang}\email[]{lmf2021@buaa.edu.cn (corresponding author)}
\affiliation{LMIB(Beihang University), Ministry of Education, and School of Mathematical Sciences, Beihang University, Beijing 100191, China}

\begin{abstract}
Quantum gravity between masses can produce entangled states in thought experiments. We extend the experiments to tripartite case and construct states equivalent to Greenberger-
Horne-Zeilinger states and W states
under stochastic local operations and classical communication. The entanglement relates to the evolution phases induced by gravitational interaction. When we involve more masses in the experiments, multipartite entangled states can be constructed in a similar way.
We measure the
degree of multipartite entanglement by calculating the geometric measure.
We describe the relationship between geometric measure and the evolution phases. It helps searching out the states
with robust entanglement.
\end{abstract}

\maketitle

\section{Introduction }
\label{sec:dis}
Due to the extreme weakness of gravity, its quantum effects are hard to detect. Recently, authors reported observing gravitational Aharonov-Bohm effect \cite{18,19}. This experiment measured the gravitational phase shift induced by a kilogram-scale source mass close to the wave packets. It convinced us about the quantum feature of gravity.
Bose et al. \cite{1} and Marletto et al. \cite{2} suggested two similar thought experiments to probe quantized gravity. Two neutral masses are separable initially, split into superposition of spatially localized states in an inhomogeneous magnetic field. The mutual gravitational interaction between components of superposition will evolute relative phases and transform initial separable state into bipartite entangled state. These quantum phases correlate with their interaction time and can be detected by entanglement witnesses.
Even though the dominant contribution of interaction is Newtonian at the low energy limits, the entanglement between two masses can verify quantum signatures of gravity. Because local operations and classical communication (LOCC) does not create entanglement, the entanglement can be generated only by non-classical mediator. There are also some discussions on Post-Newtonian order corrections in \cite{11,15}. The feasibility of the hypothetical experiments have been discussed in the original articles \cite{1,2}.

 There are some factors which may pollute the entanglement in Bose's experiment, such as Casimir-Polder forces, van der Waals forces or other electro-magnetic interactions.
By adjusting the parameters or setups of the experiments, some pollution can be reduced. A thought experiment was presented to study interaction mediated only by gravity between two hypothetical neutrino-like particles \cite{10}. A modified model of original experiment, the symmetric setup, enhanced the gravity interaction to against the noisy dynamics, such as stochastic fluctuations of the parameters or decoherence induced by environmental interaction \cite{5}. However, these studies all focused on bipartite entanglement, as far as we know, multipartite entanglement is little understood.

In this paper, we extend the quantum gravity inducing entanglement to multiqubit case.
Multiqubit entanglement plays an important role in quantum information, computation and communication. Neutral atoms' entanglement can be generated in the platforms, such as cavity-QED \cite{29}, neutral-atom tweezer arrays \cite{28}. In these platforms, arbitrarily-shaped three-dimensional arrays of atoms are realizable, which enables us to keep neutral masses in certain distances as the thought experiments acquired. So gravity interaction between the masses can generate multipartite entanglement. There are also various platforms generating multipartite entanglement with photons or ions \cite{20,21,22,23,24}. Photonic experiments entangled 14 photons to realize Greenberger-Horne-Zeilinger (GHZ) states by interleave single-photon emissions with atomic rotations \cite{20}. In a linear Paul trip, GHZ states were produced with up to 24 ions, mediated by the M\o lmer-S\o rensen gate \cite{24}.
 These attempts went a step further in quantum computation. In the future, the major problems still are how to increase the efficiency of generating entanglement and protect systems against decoherence. In \cite{30,31}, the entanglement of three and more qubits has been discussed. This work considers the case with multiple massive particles and suggests a theoretical path to produce entangled states induced by mutual gravitational interaction of neutral masses.

In multipartite system, the Hamiltonian leads to relative evolution phases with special spatial symmetry. The interaction is similar to the bipartite case in \cite{5}. In this protocol, the quantum gravity can generate GHZ states, but not W states.
We can classify equivalent entangled states under stochastic LOCC (SLOCC). They contain the same kind of entanglement and are suited to implement the same tasks of quantum information theory \cite{26}. We denote GHZ-type states as those states equivalent to GHZ states under SLOCC, and similarly for W-type states. In Theorem \ref{th:L=1}, we show that the gravitation can generate $N$-qubit GHZ-type states. However, considering the weakness of gravity, transform $N$ masses into entangled states is difficult. We suggest a way to generate $N$-qubit GHZ-type states by getting $(N-2)$-qubit GHZ-type states and Bell states entangled. The $(N-2)$-qubit GHZ-type states can be produced in the gravitational entanglement apparatus too. It provides an approach to extend existing multipartite entanglement platforms by involving more qubits and makes the experiment more feasible.
To address the degree of the gravity induced entanglement from a geometric viewpoint \cite{16,17}, we derive the geometric measure (GM) of entanglement for the tripartite case. GM quantifies the entanglement by measuring the distance between the entangled state and the nearest product state. So we can seek out the robust entangled final states induced by gravity with the results in Theorem \ref{th:L=2}. We measure the entanglement by negativity to support this results. On the other hand, the interaction time oscillation period of the measurements, which is detectable, can be a clear signal of quantum gravity.
These attempts can enrich our knowledge of quantum gravity.

The rest of this paper is constructed as follows. In Sec. \ref{sec:pre}, we introduce Bose's experiment in detail. In Sec. \ref{sec:mq}, more masses are led in to construct multiqubit entanglement and GHZ-type states in the symmetric setup. In Sec. \ref{sec:gm}, we derive the GM and negativity of entanglement for the three-qubit case. Sec. \ref{sec:con} makes conclusion and outlook.

\section{Preliminaries }
\label{sec:pre}
\begin{figure}[h!]
\centerline{
\includegraphics[width=0.8\textwidth]{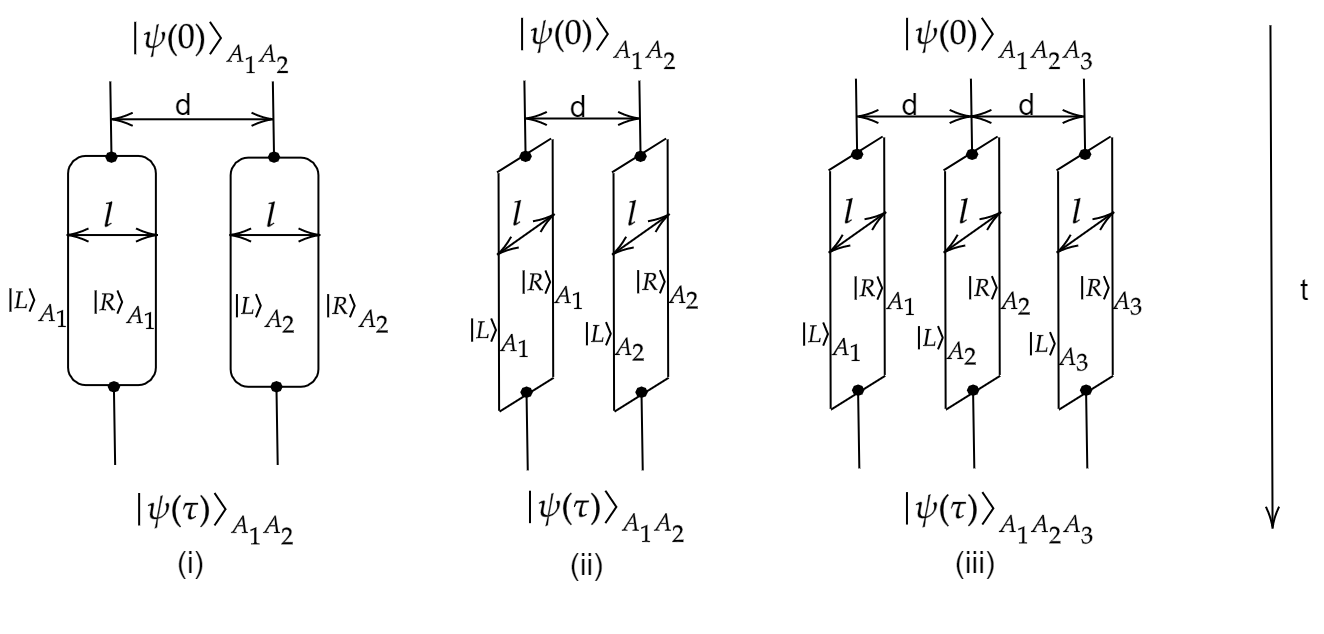}
}\caption{Two masses at distance d from each other, both split into superpositions of spatially localized states at distance $l$. (i)The original setup proposed in \cite{1}, superpositions are parallel to the initial separations. (ii) The symmetric setup proposed in \cite{5}, superpositions are orthogonal to the initial separations. (iii) The symmetric setup for three-qubit. }
\label{fig:1}
\end{figure}

Bose et al. proposed a thought experiment in \cite{1}. Two neutral masses $m_1$ and $m_2$, in an inhomogeneous magnetic field, both split into a superposition of two spatially separated states $\ket{L}$ and $\ket{R}$ for a time $\tau$. As we can see in Fig. \ref{fig:1}(i), $l$ is the distance between components of superposition, and $d$ is the distance between the centres of two masses.

Initially, two masses $A_1$ and $A_2$ are separated, each one splits into two superpositions.
\begin{equation}
\label{eq:n=initial}
   \ket{\psi(0)}_{A_1A_2}=\frac{1}{\sqrt{2}}\big(\ket{L}_{A_1}+\ket{R}_{A_1}\big)\frac{1}{\sqrt{2}}\big(\ket{L}_{A_2}+\ket{R}_{A_2}\big).
\end{equation}

The Schr{\"o}dinger equation reveals the time evolution of the state,
\begin{equation}
\label{eq:n=Sch}
   i\hbar\partial_t\psi(r,t)=\left[-\frac{\hbar ^2}{2m}\nabla^2+V(r)\right]\psi(r,t).
\end{equation}
The Hamiltonian $H=-\frac{\hbar ^2}{2m}\nabla^2+V=H_0+V$ can be separated into two parts, $H_0$ relates to the free particle's behaviour and $V$ is the mutual gravity potential.

During the interaction time, these components keep stable distances from each other, so they have time independent gravity interaction energies.
By calculating the propagator and the scattering process, the mutual gravitation potential can be described as \cite{8,12,14},
\begin{equation}
\label{eq:n=V}
   V(r)=-\frac{Gm_1m_2}{r}(1+3\frac{G(m_1+m_2)}{rc^2}+\frac{41G\hbar}{10\pi r^2c^3}).
\end{equation}
The general relativistic correction and quantum correction to Newtonian potential are extremely small. Even if we take the Newtonian approximation of potential, the entanglement can reflect the quantum nature of gravity too. Actually in this setting, gravity acts as a mediator in quantum mechanics, the entanglement will be detected only through non-classic dynamics. So in this paper, we just take the Newtonian approximation of potential for simplicity.

Since the potential is independent of time, the Schr{\"o}dinger equation for the state $\psi(r,t)$ can be solved by separating variables, $\psi(r,t)=e^{\frac{-iEt}{\hbar}}\psi(r)$, the evolution phase is related to the potential $e^{i\phi}\equiv e^{\frac{-iVt}{\hbar}}$.

Because the states $\ket{L}$ and $\ket{R}$ can be
separated by different distances (including $d, d-l, d+l$), the mutual gravity interaction can induce different rates of phase evolution in Stern-Gerlach(SG) apparatus, for simplicity, written as,
\begin{equation}
\label{eq:n=end}
   \ket{\psi(\tau)}_{A_1A_2}=\frac{1}{2}\Big[\ket{L(\tau)}_{A_1}\big(e^{i\phi_1}\ket{L(\tau)}_{A_2}+e^{i\phi_2}\ket{R(\tau)}_{A_2}\big)
   +\ket{R(\tau)}_{A_1}\big(e^{i\phi_3}\ket{L(\tau)}_{A_2}+e^{i\phi_1}\ket{R(\tau)}_{A_2}\big)\Big].
\end{equation}
If we take Newtonian approximation of the potential in Eq. (\ref{eq:n=V}), the evolution phases are \cite{1}
\begin{equation}
\label{eq:n=phi2}
\phi_1\sim \frac{Gm_1m_2\tau}{\hbar d},\qquad \phi_2\sim \frac{Gm_1m_2\tau}{\hbar (d+l)},\qquad \phi_3\sim \frac{Gm_1m_2\tau}{\hbar (d-l)}.
\end{equation}
We denote the relative phases $\Delta\phi_2=\phi_2-\phi_1$ and $\Delta \phi_3=\phi_3-\phi_1$.
The entanglement depends on the relative phases and is irrelative to common phase $\phi_1$. There is an exceptional case, when $\Delta\phi_2+\Delta\phi_3=2n\pi$, $n\in \mathbb{Z}$ ($\mathbb{Z}$ is the integer set), $\ket{\psi(\tau)}_{A_1A_2}$ is separable.

The parameters are chosen as $m_1, m_2\sim 10^{-14}kg$, $d\sim 450\mu m$, $l\sim 250\mu m$, $\tau\sim 2.5s$, the entanglement witness is  $W=X\otimes Z+Y\otimes Y$ \cite{1}. The expectation value $\left \langle W \right \rangle >1$ is the signal of entanglement.
This witness has a suboptimal detection areas and may be ineffective for small entanglement which corresponds to short interaction time $\tau$.
Nevertheless, a long interaction time is also infeasible, because when we consider the decoherence effect from earth's gravity, it is difficult to keep superposition states free falling in such a long time. So a much broader class of witnesses \cite{6,7} was suggested to detect the greatest volume of entangled states and make the setup of experiment more feasible.
 Authors constructed optimal fidelity witnesses by maximally entangled states to shorten the required interaction time, such as $W=I\otimes I-X\otimes X-Z\otimes Y-Y\otimes Z$. These witnesses are sensitive to very small entanglement, so they are valid at the beginning of free-fall. They also broaden the detection area in the space of phases.
With these instruments, the positive results will announce the quantum gravity. An entanglement witness was suggested to detect spinless entanglement between microspheres with massive spatial qubits too \cite{25}.
 However, there are also some doubts about the effectiveness of these witnesses of quantized gravity. General configuration-ensemble models and mean-field semiclassical gravity models were supposed to explain the entanglement with classical gravity \cite{3,4}.

A modified symmetric setup was suggested in \cite{5}, each mass splits into two superposition states, which are in the direction orthogonal to initial separation, see Fig. \ref{fig:1}(ii). Compared with the parallel split mode in Fig. \ref{fig:1}(i), the symmetric setup permits a reduced distance between masses, that is useful in keeping the distance constant and enhancing gravity interaction. In this case, there is only one relative phase $\Delta\phi=\frac{Gm_1m_2\tau}{\hbar}(\frac{1}{d}-\frac{1}{\sqrt{l^2+d^2}})$, the final state in Eq. (\ref{eq:n=end}) becomes
\begin{equation}
\label{eq:n=final}
 \ket{\psi(\tau)}_{A_1A_2}=\frac{e^{i\phi_1}}{2}\Big[\ket{L(\tau)}_{A_1}\big(\ket{L(\tau)}_{A_2}+e^{i\Delta\phi}\ket{R(\tau)}_{A_2}\big)
   +\ket{R(\tau)}_{A_1}\big(e^{i\Delta\phi}\ket{L(\tau)}_{A_2}+\ket{R(\tau)}_{A_2}\big)\Big].
\end{equation}

The distances of qubits in this setup display some symmetry which constraints the number of independent phases. The next sections will consider entanglement in this setup.

\section{Multiqubit Entanglement}
\label{sec:mq}
In this section, we extend gravity interaction induced entanglement to multipartite case. GHZ states and W states are the maximum entangled states under geometric entanglement metric. They play important roles in multipartite entanglement.
Besides, since the equivalence class of GHZ or W states under SLOCC contain the same kind of entanglement, we also focus on GHZ-type states and W-type states.
 Two states are equivalent under SLOCC when they are related by a local
invertible operator \cite{26}. For example, if $\ket{\psi}=(M_1 \otimes ...\otimes M_n)\ket{GHZ}$ , where each $M_i$ is an invertible matrix, then $\ket{\psi}$ is in the equivalence class of GHZ states.
We will construct GHZ-type states under the gravitational symmetric setup.
Nevertheless, if we want to construct W-type states, more freedom in parameters of the setup should be involved in. We will respectively construct three-qubit and four-qubit GHZ-type states in Sec. \ref{sec:sub1} and \ref{sec:sub2}, and then extend to N-qubit case in Sec. \ref{sec:sub3} and \ref{sec:sub4}. The final states generated in the apparatus of symmetric setup should obey some spatial symmetry, which is presented in Sec. \ref{sec:app}.

\subsection{The three-qubit case}
\label{sec:sub1}
First we introduce one more mass in Eq. (\ref{eq:n=initial}), as described in Fig. \ref{fig:1}(iii).
\begin{equation}
\label{eq:n=initial2}
   \ket{\psi(0)}_{A_1A_2A_3}=\frac{1}{\sqrt{2}}\big(\ket{L}_{A_1}+\ket{R}_{A_1}\big)\frac{1}{\sqrt{2}}\big(\ket{L}_{A_2}
   +\ket{R}_{A_2}\big)\frac{1}{\sqrt{2}}\big(\ket{L}_{A_3}+\ket{R}_{A_3}\big).
\end{equation}
The time evolution of this state is similar to the bipartite case in Eq. (\ref{eq:n=Sch}). The gravitation potential in Hamiltonian is related to the distances between three components of superpositions. We consider the symmetric setup, and there are three different evolution phases in the final state. The gravitation potential among the left components $\ket{L}_{A_1}$, $\ket{L}_{A_2}$ and $\ket{L}_{A_3}$ in Fig. \ref{fig:1}(iii) is $V=-Gm_1m_2(\frac{1}{d}+\frac{1}{d}+\frac{1}{2d})$, so does the potential among the right components. The potential among $\ket{L}_{A_1}$, $\ket{R}_{A_2}$ and $\ket{L}_{A_3}$ is $V=-Gm_1m_2(\frac{1}{2d}+\frac{1}{\sqrt{d^2+l^2}}+\frac{1}{\sqrt{d^2+l^2}})$, so does the potential among $\ket{R}_{A_1}$, $\ket{L}_{A_2}$ and $\ket{R}_{A_3}$. The potential among $\ket{L}_{A_1}$, $\ket{L}_{A_2}$ and $\ket{R}_{A_3}$ is $V=-Gm_1m_2(\frac{1}{d}+\frac{1}{\sqrt{4d^2+l^2}}+\frac{1}{\sqrt{d^2+l^2}})$, so do rest parts in the state. The final state can be written as
\begin{equation}
\label{eq:n=three}
\begin{aligned}
\ket{\psi(\tau)}_{A_1A_2A_3} & =\frac{1}{2\sqrt{2}}\Big[e^{i \varphi_1}\big(\ket{L(\tau)}_{A_1}\ket{L(\tau)}_{A_2}\ket{L(\tau)}_{A_3}+\ket{R(\tau)}_{A_1}\ket{R(\tau)}_{A_2}\ket{R(\tau)}_{A_3}\big)\\
  & +e^{i\varphi_2}\big(\ket{L(\tau)}_{A_1}\ket{L(\tau)}_{A_2}\ket{R(\tau)}_{A_3}
+\ket{L(\tau)}_{A_1}\ket{R(\tau)}_{A_2}\ket{R(\tau)}_{A_3} \\
   & +\ket{R(\tau)}_{A_1}\ket{L(\tau)}_{A_2}\ket{L(\tau)}_{A_3} +\ket{R(\tau)}_{A_1}\ket{R(\tau)}_{A_2}\ket{L(\tau)}_{A_3}\big)\\
   &  +e^{i \varphi_3}\big(\ket{L(\tau)}_{A_1}\ket{R(\tau)}_{A_2}\ket{L(\tau)}_{A_3}+\ket{R(\tau)}_{A_1}\ket{L(\tau)}_{A_2}\ket{R(\tau)}_{A_3}\big)\Big].
\end{aligned}
\end{equation}
The evolution phases are
\begin{equation}
\label{eq:n=varphi}
\begin{aligned}
 & \varphi_1\sim \frac{5Gm_1m_2\tau}{2\hbar d},\\
 & \varphi_2\sim \frac{Gm_1m_2\tau}{\hbar}(\frac{1}{d}+\frac{1}{\sqrt{4d^2+l^2}}+\frac{1}{\sqrt{d^2+l^2}}),\\
  & \varphi_3\sim \frac{Gm_1m_2\tau}{\hbar}(\frac{1}{2d}+\frac{2}{\sqrt{d^2+l^2}}).
\end{aligned}
\end{equation}

For simplicity, we denote $\ket{L(\tau)}$ as $\ket{0}$, $\ket{R(\tau)}$ as $\ket{1}$, $\Delta\varphi_2=\varphi_2-\varphi_1$, $\Delta\varphi_3=\varphi_3-\varphi_1$.
Eq. (\ref{eq:n=three}) can be written as
\begin{equation}
\label{eq:n=three2}
\begin{split}
 \ket{\psi}=\frac{e^{i\varphi_1}}{2\sqrt{2}}\Big[\ket{000}+\ket{111}
   +e^{i\Delta\varphi_2}\big(\ket{001}+\ket{011}+
   \ket{100}+\ket{110}\big)
   +e^{i\Delta\varphi_3}\big(\ket{010}+\ket{101}\big)\Big].
\end{split}
\end{equation}

The entanglement of Eq. (\ref{eq:n=three2}) comes from the relative evolution phases $\Delta\varphi_2$ and $\Delta\varphi_3$. In some certain situation, the state in Eq. (\ref{eq:n=three2}) may become the separable state, GHZ state or GHZ-type state.
In the following, we will discuss the three cases (i), (ii) and (iii).

(i) The state in Eq. (\ref{eq:n=three2}) is a separable state. We write it as
\begin{equation}
\label{eq:n=three8}
\begin{aligned}
 \ket{\psi}=\frac{e^{i\varphi_1}}{2\sqrt{2}} & \bigg\{\ket{0}_{A_1}\Big[\ket{00}_{A_2A_3}
   +e^{i\Delta\varphi_2}\big(\ket{01}_{A_2A_3}+\ket{11}_{A_2A_3}\big)
   +e^{i\Delta\varphi_3}\ket{10}_{A_2A_3}\Big]\\
    & +\ket{1}_{A_1}\Big[\ket{11}_{A_2A_3}+e^{i\Delta\varphi_2}\big(\ket{00}_{A_2A_3}+\ket{10}_{A_2A_3}\big)
   +e^{i\Delta\varphi_3}\ket{01}_{A_2A_3}\Big]\bigg\}.
\end{aligned}
\end{equation}
So in the range space of the reduced density operator of the second and third qubits, there are two vectors,
\begin{equation}
\label{eq:n=three13}
\begin{aligned}
\ket{a} & =\frac{1}{2}\big[\ket{00}_{A_2A_3}
   +e^{i\Delta\varphi_2}\big(\ket{01}_{A_2A_3}+\ket{11}_{A_2A_3}\big)
   +e^{i\Delta\varphi_3}\ket{10}_{A_2A_3}\big],\\
   \ket{b} & =\frac{1}{2}\big[\ket{11}_{A_2A_3}+e^{i\Delta\varphi_2}\big(\ket{00}_{A_2A_3}+\ket{10}_{A_2A_3}\big)
   +e^{i\Delta\varphi_3}\ket{01}_{A_2A_3}\big].
\end{aligned}
\end{equation}
If the vectors $\ket{a}$ and $\ket{b}$ are linearly dependent, the final state $\ket{\psi}$ is a separable state.
For example, by choosing $\Delta\varphi_3=2n\pi$, $n\in \mathbb{Z}$, the state in Eq. (\ref{eq:n=three2}) is separable,
\begin{equation}
\label{eq:n=three6}
\begin{split}
 \ket{\psi_1}=\frac{e^{i\varphi_1}}{2\sqrt{2}}[\ket{0}_{A_1}\otimes(\ket{0}+\ket{1})_{A_2}\otimes(\ket{0}+e^{i\Delta\varphi_2}\ket{1})_{A_3}
 +\ket{1}_{A_1}\otimes(\ket{0}+\ket{1})_{A_2}\otimes(e^{i\Delta\varphi_2}\ket{0}+\ket{1})_{A_3}].
\end{split}
\end{equation}
The state in Eq. (\ref{eq:n=three6}) will be fully separable states for $\Delta\varphi_2=n\pi$.
\begin{equation}
\label{eq:n=three3}
\begin{split}
 \ket{\psi_2}=e^{i\varphi_1}(\frac{\ket{0}\pm\ket{1}}{\sqrt{2}})_{A_1}\otimes(\frac{\ket{0}\pm\ket{1}}{\sqrt{2}})_{A_2}\otimes(\frac{\ket{0}\pm\ket{1}}{\sqrt{2}})_{A_3}.
\end{split}
\end{equation}

(ii) The state in Eq. (\ref{eq:n=three2}) is a GHZ state.

When $e^{i\Delta\varphi_3}\neq 1$, the state in  Eq. (\ref{eq:n=three2}) becomes a genuinely entangled state.
The reduced density operator of the first, second and third qubit respectively is
\begin{equation}
\label{eq:n=ghz}
\begin{aligned}
 & \rho_{A_1}=\rho_{A_3}=\frac{1}{8}\big[4(\ket{0}\bra{0}+\ket{1}\bra{1})+(e^{i\Delta\varphi_2}+e^{-i\Delta\varphi_2}
+e^{i(\Delta\varphi_2-\Delta\varphi_3)}+e^{i(\Delta\varphi_3-\Delta\varphi_2)})(\ket{0}\bra{1}+\ket{1}\bra{0})\big],\\
 & \rho_{A_2}=\frac{1}{8}\big[4(\ket{0}\bra{0}+\ket{1}\bra{1})+(e^{i\Delta\varphi_3}+e^{-i\Delta\varphi_3}+2)(\ket{0}\bra{1}+\ket{1}\bra{0})\big].
\end{aligned}
\end{equation}
When $e^{i\Delta\varphi_2}+e^{-i\Delta\varphi_2}
+e^{i(\Delta\varphi_2-\Delta\varphi_3)}+e^{i(\Delta\varphi_3-\Delta\varphi_2)}=0$ and $e^{i\Delta\varphi_3}+e^{-i\Delta\varphi_3}+2=0$, the state in Eq. (\ref{eq:n=three2}) is a GHZ state. That means $\Delta\varphi_3=(2n+1)\pi$.
For example, by choosing $e^{i\Delta\varphi_2}=i$, $e^{i\Delta\varphi_3}=-1$. Eq. (\ref{eq:n=three2}) becomes
\begin{equation}
\label{eq:n=ghz2}
\begin{split}
\ket{\psi_3}=\frac{e^{i\varphi_1}}{2\sqrt{2}}[(\ket{0}+i\ket{1})_{A_1}\otimes\ket{0}_{A_2}\otimes(\ket{0}+i\ket{1})_{A_3}
+(i\ket{0}+\ket{1})_{A_1}\otimes\ket{1}_{A_2}\otimes(i\ket{0}+\ket{1})_{A_3}].
\end{split}
\end{equation}
The state in Eq. (\ref{eq:n=ghz2}) is a GHZ state, because $\ket{\psi}=(\frac{\sigma_z+\sigma_y}{2})_{A_1}\otimes I_{A_2}\otimes(\frac{\sigma_z+\sigma_y}{2})_{A_3}(\ket{000}+\ket{111})$.

When we choose $e^{i\Delta\varphi_2}=1$ and $e^{i\Delta\varphi_3}=-1$, the state in Eq. (\ref{eq:n=three2}) is LU equivalent to a GHZ state,
\begin{equation}
\label{eq:n=ghz3}
\begin{split}
 \ket{\psi_4}=\frac{e^{i\varphi_1}}{2\sqrt{2}}[(\ket{00}+\ket{11})_{A_1A_2}\otimes(\ket{0}+\ket{1})_{A_3}
 +(\ket{10}-\ket{01})_{A_1A_2}\otimes(\ket{0}-\ket{1})_{A_3}].
\end{split}
\end{equation}

(iii) The state in Eq. (\ref{eq:n=three2}) is a GHZ-type state.

The state in Eq. (\ref{eq:n=three2}) becomes a GHZ-type state when there exist two linearly independent product vectors in the range space of the reduced density operator.
That means if the vectors $\ket{a}$ and $\ket{b}$ construct a product state $\ket{\psi^p}=\ket{a}+x\ket{b}$,
there exist two roots of $x$ which satisfy the equation
\begin{equation}
\label{eq:n=x}
\begin{aligned}
  & (1+e^{i\Delta\varphi_2}x)(e^{i\Delta\varphi_2}+x)=(e^{i\Delta\varphi_2}+e^{i\Delta\varphi_3}x)(e^{i\Delta\varphi_3}+e^{i\Delta\varphi_2}x).\\
\end{aligned}
\end{equation}
For example,
when we choose $e^{i\Delta\varphi_2}=1$ and $e^{i\Delta\varphi_3}=i$, the state in Eq. (\ref{eq:n=three2}) becomes GHZ-type states,
\begin{equation}
\label{eq:n=three5}
\begin{split}
 \ket{\psi_5}=\frac{e^{i\varphi_1}}{2\sqrt{2}}[(\ket{00}+\ket{11})_{A_1A_2}\otimes(\ket{0}+\ket{1})_{A_3}
 +\ket{10}_{A_1A_2}\otimes(\ket{0}+i\ket{1})_{A_3}+\ket{01}_{A_1A_2}\otimes(i\ket{0}+\ket{1})_{A_3}].
\end{split}
\end{equation}
We can see that, there are greater chances for Eq. (\ref{eq:n=three2}) be a GHZ-type state than be a GHZ state. So we will consider GHZ-type states in next sections only.

On the other hand if there exists a multiple root for Eq. (\ref{eq:n=x}), the state in  Eq. (\ref{eq:n=three2}) becomes a W-type state.
For W-type states, the phases should satisfy $(1+e^{i\Delta\varphi_3})^2=4e^{2i\Delta\varphi_2}$. That means $e^{i\Delta\varphi_3}=1$, though in this case $\ket{\psi_1}$ is separable, actually there is no solution for the equation.
So under this setup, we cannot produce W-type states.

\subsection{The four-qubit case}
\label{sec:sub2}
In this part, we extend the modes of Fig. \ref{fig:1}(iii) to the four-qubit case.
The final state can be written as
\begin{equation}
\label{eq:n=four1}
\begin{aligned}
 \ket{\psi}= & \frac{1}{4}\Big[e^{i\varphi'_1}\big(\ket{0000}+\ket{1111}\big)
   +e^{i\varphi'_2}\big(\ket{0001}+\ket{0111}+
   \ket{1000}+\ket{1110}\big)\\
   &  +e^{i\varphi'_3}\big(\ket{0010}+\ket{1101}+\ket{0100}+\ket{1011}\big)
   +e^{i\varphi'_4}\big(\ket{1100}+\ket{0011}\big)\\
    & +e^{i\varphi'_5}\big(\ket{1010}+\ket{0101}\big) +e^{i\varphi'_6}\big(\ket{1001}+\ket{0110}\big)\Big].
\end{aligned}
\end{equation}
The evolution phases are similar to Eq. (\ref{eq:n=varphi}),
\begin{equation}
\label{eq:n=varphi2}
\begin{aligned}
\varphi'_1\sim  & \frac{13Gm_1m_2\tau}{3\hbar d},\\
\varphi'_2\sim  & \frac{Gm_1m_2\tau}{\hbar}(\frac{5}{2d}+\frac{1}{\sqrt{9d^2+l^2}}+\frac{1}{\sqrt{4d^2+l^2}}+\frac{1}{\sqrt{d^2+l^2}}),\\
 \varphi'_3\sim  & \frac{Gm_1m_2\tau}{\hbar}(\frac{11}{6d}+\frac{1}{\sqrt{4d^2+l^2}}+\frac{2}{\sqrt{d^2+l^2}}),\\
 \varphi'_4\sim  & \frac{Gm_1m_2\tau}{\hbar}(\frac{2}{d}+\frac{1}{\sqrt{9d^2+l^2}}+\frac{2}{\sqrt{4d^2+l^2}}+\frac{1}{\sqrt{d^2+l^2}}),\\
 \varphi'_5\sim  & \frac{Gm_1m_2\tau}{\hbar}(\frac{1}{d}+\frac{1}{\sqrt{9d^2+l^2}}+\frac{3}{\sqrt{d^2+l^2}}),\\
 \varphi'_6\sim  & \frac{Gm_1m_2\tau}{\hbar}(\frac{4}{3d}+\frac{2}{\sqrt{4d^2+l^2}}+\frac{2}{\sqrt{d^2+l^2}}).
\end{aligned}
\end{equation}
When we extract $e^{i\varphi'_1}$, the relative phases are $\Delta\varphi'_i=\varphi'_i-\varphi'_1$.
If the relative phases are chosen as $e^{i\Delta\varphi'_2}=e^{i\Delta\varphi'_3}$ and $e^{i\Delta\varphi'_4}=e^{i\Delta\varphi'_5}=e^{i\Delta\varphi'_6}=1$, the state will be GHZ-type.
For example, if we choose $e^{i\Delta\varphi'_2}=e^{i\Delta\varphi'_3}=i$, then Eq. (\ref{eq:n=four1}) becomes
\begin{equation}
\label{eq:n=four2}
\begin{aligned}
 \ket{\psi}=  & \frac{1}{4}e^{i\varphi'_1}\Big\{(\ket{00}+\ket{11})_{A_1A_2}
   \big[\ket{0}_{A_3}(\ket{0}+i\ket{1})_{A_4}+\ket{1}_{A_3}(i\ket{0}+\ket{1})_{A_4}\big]\\
    & +(\ket{10}+\ket{01})_{A_1A_2}
   \big[\ket{0}_{A_3}(i\ket{0}+\ket{1})_{A_4}+\ket{1}_{A_3}(\ket{0}+i\ket{1})_{A_4}\big]\Big\}.
\end{aligned}
\end{equation}

Looking through the three-qubit and the four-qubit cases, we can see that the relative phases are related to the distances of qubits.
As in Fig. \ref{fig:1}(iii), the distances are constrained by some spatial symmetry. We will present the details in Sec. \ref{sec:app}.

When we extend the results to $N$-qubit case, we have Theorem \ref{th:L=1}.
\begin{theorem}
\label{th:L=1}	
When masses are split into superpositions in symmetric setup as Fig. \ref{fig:1}(iii) described,
the gravity interaction between superpositions can produce N-qubit GHZ-type entangled states.
\end{theorem}

Since there is some difference between the $(2N+1)$-qubit case and the $2N$-qubit case, we will show Theorem \ref{th:L=1} from two aspects in Sec. \ref{sec:sub3} and \ref{sec:sub4} . First, let us consider the $(2N+1)$-qubit case.

\subsection{The $(2N+1)$-qubit case}
\label{sec:sub3}
We define a one-qubit basis,
\begin{equation}
\label{eq:n=bases}
 \ket{m_+}=\frac{\ket{0}+\ket{1}}{\sqrt{2}}, \qquad      \ket{m_-}=\frac{\ket{0}-\ket{1}}{\sqrt{2}}.
\end{equation}
They are linearly independent,
then we can construct a three-qubit state like this,
\begin{equation}
\label{eq:n=3qubits}
\begin{aligned}
 \ket{\psi^3}  & =\frac{\ket{00}+\ket{11}}{2}\ket{m_+}+\frac{\ket{10}-\ket{01}}{2}\ket{m_-}\\
   & =\frac{1}{2\sqrt{2}}\Big[\ket{0}\big(\ket{00}+\ket{11}+\ket{01}-\ket{10}\big)+\ket{1}\big(\ket{00}+\ket{11}+\ket{10}-\ket{01}\big)\Big].
\end{aligned}
\end{equation}
Obviously $\ket{c}=\frac{1}{\sqrt{2}}[\ket{00}+\ket{11}+\ket{01}-\ket{10}]$ and $\ket{d}=\frac{1}{\sqrt{2}}[\ket{00}+\ket{11}+\ket{10}-\ket{01}]$ are linearly independent, and
they can construct two separable states
\begin{equation}
\label{eq:n=otherbases}
\begin{split}
\ket{c}+i\ket{d} & =\frac{1+i}{\sqrt{2}}(\ket{0}+i\ket{1})(\ket{0}-i\ket{1}),\\
\ket{c}-i\ket{d} & =\frac{1-i}{\sqrt{2}}(\ket{0}-i\ket{1})(\ket{0}+i\ket{1}).
\end{split}
\end{equation}
They span bipartite Hilbert space, so the state in Eq. (\ref{eq:n=3qubits}) is a three-qubit GHZ-type state, just be the same as Eq. (\ref{eq:n=ghz3}).

Next, we introduce a back-up basis
\begin{equation}
\label{eq:n=3bases}
\begin{split}
\ket{\psi'^3}=\frac{\ket{00}+\ket{11}}{2}\ket{m_-}-\frac{\ket{10}-\ket{01}}{2}\ket{m_+}.
\end{split}
\end{equation}
Since $\ket{\psi'^3}=(\sigma_x)_{A_3}(-\sigma_z)_{A_3}\ket{\psi^3}$, it is also a GHZ-type state, and is linearly independent with $\ket{\psi^3}$. The states in $\ket{\psi^3}$ and $\ket{\psi'^3}$ construct a three-qubit basis. We should notice that the state in $\ket{\psi'^3}$ is not that kind final state produced in our apparatus, because it doesn't satisfy the spatial symmetry described in Theorem \ref{th:L=3}.

Now we can construct a five-qubit GHZ-type state by three-qubit basis,
\begin{equation}
\label{eq:n=five}
\begin{split}
 \ket{\psi^5}=\frac{\ket{00}+\ket{11}}{2}\ket{\psi^3}+\frac{\ket{10}-\ket{01}}{2}\ket{\psi'^3}.
\end{split}
\end{equation}
It can be written as
\begin{equation}
\label{eq:n=five2}
\begin{aligned}
\ket{\psi^5}= & \frac{1}{4}\bigg\{\Big[\big(\ket{00}+\ket{11}\big)+i\big(\ket{01}-\ket{10}\big)\Big]\big(\ket{\psi^3}+i\ket{\psi'^3}\big)\\
  & +\Big[\big(\ket{00}+\ket{11}\big)-i\big(\ket{01}-\ket{10}\big)\Big]\big(\ket{\psi^3}-i\ket{\psi'^3}\big)\bigg\}.
\end{aligned}
\end{equation}
So it is a GHZ-type state.
The five-qubit back-up basis is
\begin{equation}
\label{eq:n=5bases}
\ket{\psi'^5}=\frac{\ket{00}+\ket{11}}{2}\ket{\psi'^3}-\frac{\ket{10}-\ket{01}}{2}\ket{\psi^3}.
\end{equation}
It has the same feature as $\ket{\psi'^3}$.
Then we can construct a seven-qubit GHZ-type state in the same way.
So generally, a $(2N+1)$-qubit GHZ-type state can be presented as
\begin{equation}
\label{eq:n=2n1}
\begin{aligned}
 \ket{\psi^{2N+1}} & =\frac{\ket{00}+\ket{11}}{2}\ket{\psi^{2N-1}}+\frac{\ket{10}-\ket{01}}{2}\ket{\psi'^{2N-1}}\\
 & =\frac{1}{4}\bigg\{\Big[\big(\ket{00}+\ket{11}\big)+i\big(\ket{01}-\ket{10}\big)\Big]\big(\ket{\psi^{2N-1}}+i\ket{\psi'^{2N-1}}\big)\\
  &  \quad +\Big[\big(\ket{00}+\ket{11}\big)-i\big(\ket{01}-\ket{10}\big)\Big]\big(\ket{\psi^{2N-1}}-i\ket{\psi'^{2N-1}}\big)\bigg\}.
\end{aligned}
\end{equation}

As the criterion proposed in Theorem \ref{th:L=3} shown, the $(2N+1)$-qubit GHZ-type state $\ket{\psi^{2N+1}}$ can be produced in the symmetric setup, but the back-up basis $\ket{\psi'^{2N+1}}$ can not be generated in the apparatus.

\subsection{The $2N$-qubit case}
\label{sec:sub4}
The $2N$-qubit case is similar as the $(2N+1)$-qubit case, the two-qubit basis is
\begin{equation}
\label{eq:n=bases2}
 \ket{k_+}=\frac{1}{2}\Big[\big(\ket{00}+\ket{11}\big)+i\big(\ket{01}+\ket{10}\big)\Big], \qquad      \ket{k_-}=\frac{1}{2}\Big[i\big(\ket{00}+\ket{11}\big)+\big(\ket{01}+\ket{10}\big)\Big].
\end{equation}
The four-qubit GHZ-type state is
\begin{equation}
\label{eq:n=4qubits}
 \ket{\psi^4}=\frac{\ket{00}+\ket{11}}{2}\ket{k_+}+\frac{\ket{10}+\ket{01}}{2}\ket{k_-},
\end{equation}
just the same as Eq. (\ref{eq:n=four2}).

Next, we introduce the four-qubit back-up basis
\begin{equation}
\label{eq:n=4bases}
\ket{\psi'^4}=\big(\ket{00}+\ket{11}\big)\ket{k_-}+\big(\ket{10}+\ket{01}\big)\ket{k_+}=(\sigma_x)_4\ket{\psi^4}.
\end{equation}
It is GHZ-type and linearly independent with $\ket{\psi^4}$. Different from $\ket{\psi'^3}$, the state in $\ket{\psi'^4}$ can be produced in the symmetric setup.

Then we can construct a six-qubit GHZ-type state by four-qubit basis,
\begin{equation}
\label{eq:n=six}
\begin{aligned}
 \ket{\psi^6}= & \frac{\ket{00}+\ket{11}}{2}\ket{\psi^4}+\frac{\ket{10}+\ket{01}}{2}\ket{\psi'^4}\\
 = & \frac{1}{4}\bigg\{\Big[\big(\ket{00}+\ket{11}\big)-\big(\ket{10}+\ket{01}\big)\Big]\big(\ket{\psi^4}-\ket{\psi'^4}\big)\\
 & +\Big[\big(\ket{00}+\ket{11}\big)+\big(\ket{10}+\ket{01}\big)\Big]\big(\ket{\psi^4}+\ket{\psi'^4}\big)\bigg\}.
\end{aligned}
\end{equation}
The six-qubit back-up basis is
\begin{equation}
\label{eq:n=6bases}
\ket{\psi'^6}=\frac{\ket{00}+\ket{11}}{2}\ket{\psi'^4}+\frac{\ket{10}+\ket{01}}{2}\ket{\psi^4}.
\end{equation}
Generally, $2N$-qubit GHZ-type states can be presented as
\begin{equation}
\label{eq:n=2n}
\begin{aligned}
 \ket{\psi^{2N}} & =\frac{\ket{00}+\ket{11}}{2}\ket{\psi^{2N-2}}+\frac{\ket{10}+\ket{01}}{2}\ket{\psi'^{2N-2}}\\
&  =\frac{1}{4}\bigg\{\Big[\big(\ket{00}+\ket{11}\big)-\big(\ket{10}+\ket{01}\big)\Big]\big(\ket{\psi^{2N-2}}-\ket{\psi'^{2N-2}}\big)\\
 & \quad +\Big[\big(\ket{00}+\ket{11}\big)+\big(\ket{10}+\ket{01}\big)\Big]\big(\ket{\psi^{2N-2}}+\ket{\psi'^{2N-2}}\big)\bigg\}.
\end{aligned}
\end{equation}

Different from the $(2N+1)$-qubit case, the states in $\ket{\psi^{2N}}$ and $\ket{\psi'^{2N}}$ both can be produced in the symmetric setup (details in Sec. \ref{sec:app}).
It provides a new approach to generate multiqubit entangled states. By getting $(2N-2)$-qubit basis and Bell states entangled, we can obtain $2N$-qubit GHZ-type state $\ket{\psi^{2N}}$. That means if we have produced $(2N-2)$-qubit GHZ-type states by gravity interaction, entangled states with more qubits are available.
This approach can be applied in improving existing multipartite entangled platforms. If we get rid of the restrict from spatial symmetry, the approach is also feasible for the $(2N+1)$-qubit case.

\subsection{The spatial symmetry}
\label{sec:app}
In this subsection, we talked about the spatial symmetry which has been mentioned in Sec. \ref{sec:sub2}. As we can see in Sec. \ref{sec:pre}, the masses split into two superpositions, components $\ket{L}$ and $\ket{R}$ keep stable distances from each other in the apparatus of symmetric setup (as Fig. \ref{fig:1}(ii, iii) described).
The mutual gravity interaction can induce different rates of phase evolution in the final state.

\begin{figure}[h!]
\begin{minipage}{0.2\linewidth}
  \centering
        \centerline{
        \includegraphics[width=1\textwidth]{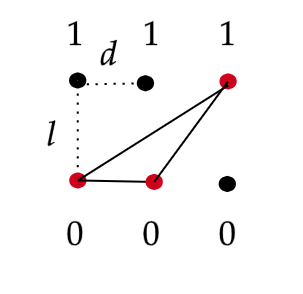}
        }
        \centerline{(i) $\ket{001}$}
    \end{minipage}
    \begin{minipage}{0.2\linewidth}
        \centering
        \centerline{
        \includegraphics[width=1\textwidth]{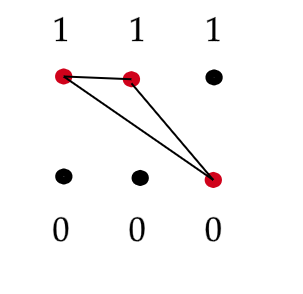}
        }
        \centerline{(ii) $\ket{110}$}
    \end{minipage}
    \begin{minipage}{0.2\linewidth}
  \centering
        \centerline{
        \includegraphics[width=1\textwidth]{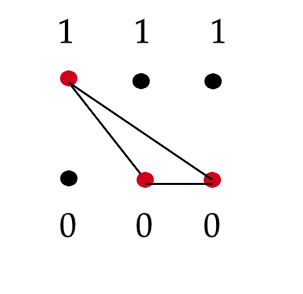}
        }
        \centerline{(iii) $\ket{100}$}
    \end{minipage}
    \begin{minipage}{0.2\linewidth}
        \centering
        \centerline{
        \includegraphics[width=1\textwidth]{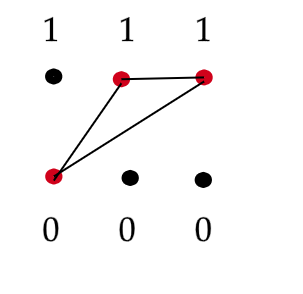}
        }
        \centerline{(iv) $\ket{011}$}
    \end{minipage}
\caption{Vertical view of Fig. \ref{fig:1}(iii). The dots stand for the superpositions of the first ($A_1$), second ($A_2$) and third ($A_3$) masses from left to right. The lines refer to the distances between components of the masses.}
\label{fig:4}
\end{figure}

In Eq. (\ref{eq:n=varphi}) and Eq. (\ref{eq:n=varphi2}), these phases are decided by the distances between $\ket{L}$ and $\ket{R}$. For example, for the three-qubit case in Eq. (\ref{eq:n=three2}), the evolution phase of $\ket{001}$ is $\varphi_2$ in Eq. (\ref{eq:n=varphi}). It relates to the distances, including $d$ between $\ket{0}_{A_1}$ and $\ket{0}_{A_2}$, $\sqrt{d^2+l^2}$ between $\ket{0}_{A_2}$ and $\ket{1}_{A_3}$, $\sqrt{4d^2+l^2}$ between $\ket{0}_{A_1}$ and $\ket{1}_{A_3}$, as Fig. \ref{fig:4}(i) described. We check Fig. \ref{fig:4} (ii, iii, iv), and find that $\ket{110}$, $\ket{100}$ and $\ket{011}$ should have the same phase $\varphi_2$, because the sum of distances in these cases are equivalent to Fig. \ref{fig:4}(i).

Other parts in Eq. (\ref{eq:n=three2}) reveal the same feature, $\ket{000}$ and $\ket{111}$ share the same phase $\varphi_1$, $\ket{010}$ and $\ket{101}$ share the same phase $\varphi_3$. However, $\varphi_1$, $\varphi_2$, $\varphi_3$ are independent. That is the spatial symmetry we talked about in Sec. \ref{sec:mq}.

\begin{figure}[h!]
\begin{minipage}{0.33\linewidth}
  \centering
        \centerline{
        \includegraphics[width=0.7\textwidth]{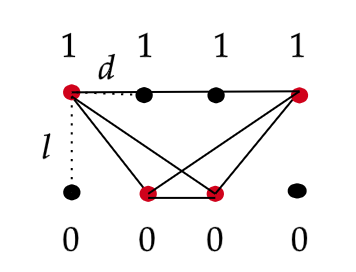}
        }
        \centerline{(i) $\ket{1001}$}
    \end{minipage}
    \begin{minipage}{0.33\linewidth}
        \centering
        \centerline{
        \includegraphics[width=0.7\textwidth]{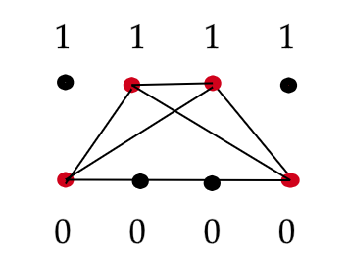}
        }
        \centerline{(ii) $\ket{0110}$}
    \end{minipage}
  \caption{In the four-qubit case, $\ket{1100}$ and $\ket{0011}$ have the same phase since they share the symmetric distances.}
\label{fig:5}
\end{figure}

For the four-qubit case in Eq. (\ref{eq:n=four1}), we classify the state into six groups with independent phases, $\big(\ket{0000}, \ket{1111}\big)$, $\big(\ket{0001}, \ket{0111},\ket{1000}, \ket{1110}\big)$,
$\big(\ket{0010}, \ket{1101}, \ket{0100}, \ket{1011}\big)$, $\big(\ket{1100}, \ket{0011}\big)$, $\big(\ket{1010}, \ket{0101}\big)$, $\big(\ket{1001}, \ket{0110}\big)$, each group of states in one bracket have the same phase. The states in the last bracket are presented in Fig. \ref{fig:5} for example. We find that the states in one bracket are symmetric if we exchange $\ket{0}$ and $\ket{1}$ or turn over the qubit.

We define some notations to help understanding. First, we denote $\widetilde{\ket{\psi}}$ as the invert state of $\ket{\psi}$, that means $\widetilde{\ket{0}}=\ket{1}$, $\widetilde{\ket{1}}=\ket{0}$ for each qubit, for example if $\ket{\psi}=\ket{001}$ then $\widetilde{\ket{\psi}}=\ket{110}$.
Next, if we turn over the qubit in the state $\ket{\psi}$, it becomes the turn over state $\ket{\psi}^{To}$, such as $(\ket{0}_A\ket{0}_B\ket{1}_C)^{To}=\ket{1}_A\ket{0}_B\ket{0}_C$. For example if $\ket{\psi}=\ket{101001}+\ket{110001}$, the turn over state $\ket{\psi}^{To}=\ket{100101}+\ket{100011}$.

We describe the spatial symmetry in Theorem \ref{th:L=3}.
\begin{theorem}
\label{th:L=3}	
In the symmetric setup described by Fig. \ref{fig:1} (ii, iii), the multipartite entangled state $\ket{\psi}$ should obey spatial symmetry that each symmetric groups in the final state should share the same evolution phase. It equals to the statements $\widetilde{\ket{\psi}}=\ket{\psi}$ and $\ket{\psi}^{To}=\ket{\psi}$.
\end{theorem}

The states we have constructed in Eq. (\ref{eq:n=2n1}) and Eq. (\ref{eq:n=2n}) under symmetric setup should follow Theorem \ref{th:L=3}.
We will check whether they satisfy above statements from Eq. (\ref{eq:n=bases}) to Eq. (\ref{eq:n=2n}).

~\\
  \indent\textbf{1.The $(2N+1)$-qubit case}

Obviously, the basis in Eq. (\ref{eq:n=bases}) satisfy,
\begin{equation}
\label{eq:n=checkpsi1}
\begin{split}
\widetilde{\ket{m_+}}=\frac{\ket{1}+\ket{0}}{\sqrt{2}}=\ket{m_+}, \qquad  \ket{m_+}^{To}=\ket{m_+},\\
\widetilde{\ket{m_-}}=\frac{\ket{1}-\ket{0}}{\sqrt{2}}=-\ket{m_-}, \qquad \ket{m_-}^{To}=\ket{m_-}.
\end{split}
\end{equation}
For three-qubit basis in Eq. (\ref{eq:n=3qubits}) and Eq. (\ref{eq:n=3bases}), the invert states are
\begin{equation}
\label{eq:n=checkpsi34}
\begin{split}
\widetilde{\ket{\psi^3}} & =\frac{\ket{11}+\ket{00}}{2}\widetilde{\ket{m_+}}+\frac{\ket{01}-\ket{10}}{2}\widetilde{\ket{m_-}}=\ket{\psi^3},\\
\widetilde{\ket{\psi'^3}} & =\frac{\ket{11}+\ket{00}}{2}\widetilde{\ket{m_-}}-\frac{\ket{01}-\ket{10}}{2}\widetilde{\ket{m_+}}=-\ket{\psi'^3}.
\end{split}
\end{equation}
 The turn over state of $\ket{\psi^3}$ is
\begin{equation}
\label{eq:n=checkpsi32}
\begin{aligned}
\ket{\psi^3}^{To} & =\ket{m_+}^{To}\frac{\ket{00}+\ket{11}}{2}+\ket{m_-}^{To}\frac{\ket{01}-\ket{10}}{2}\\
 & =\frac{1}{2\sqrt{2}}\Big[\big(\ket{0}+\ket{1}\big)\big(\ket{00}+\ket{11}\big)+\big(\ket{0}-\ket{1}\big)\big(\ket{01}-\ket{10}\big)\Big]\\
 & =\frac{1}{2\sqrt{2}}\Big[\ket{0}\big(\ket{00}+\ket{01}\big)+\ket{1}\big(\ket{11}+\ket{10}\big)+\ket{0}\big(\ket{11}-\ket{10}\big)+\ket{1}\big(\ket{00}-\ket{01}\big)\Big]\\
 & =\frac{1}{2\sqrt{2}}\Big[\ket{00}\big(\ket{0}+\ket{1}\big)+\ket{11}\big(\ket{1}+\ket{0}\big)+\ket{01}\big(\ket{1}-\ket{0}\big)+\ket{10}\big(\ket{0}-\ket{1}\big)\Big]\\
 & =\frac{1}{2\sqrt{2}}\Big[\big(\ket{00}+\ket{11}\big)\big(\ket{0}+\ket{1}\big)+\big(\ket{10}-\ket{01}\big)\big(\ket{0}-\ket{1}\big)\Big]=\ket{\psi^3}.
\end{aligned}
\end{equation}
Similar, we have
\begin{equation}
\label{eq:n=checkpsi33}
\ket{\psi'^3}^{To}=\ket{m_-}^{To}\frac{\ket{00}+\ket{11}}{2}-\ket{m_+}^{To}\frac{\ket{01}-\ket{10}}{2}=\ket{\psi'^3}.
\end{equation}

For five-qubit basis in Eq. (\ref{eq:n=five}) and Eq. (\ref{eq:n=5bases}), the invert states are
\begin{equation}
\label{eq:n=checkpsi5}
\begin{aligned}
 & \widetilde{\ket{\psi^5}}=\frac{\ket{11}+\ket{00}}{2}\widetilde{\ket{\psi^3}}+\frac{\ket{01}-\ket{10}}{2}\widetilde{\ket{\psi'^3}}=\ket{\psi^5},\\
 & \widetilde{\ket{\psi'^5}}=\frac{\ket{11}+\ket{00}}{2}\widetilde{\ket{\psi'^3}}-\frac{\ket{01}-\ket{10}}{2}\widetilde{\ket{\psi^3}}=-\ket{\psi'^5}.
\end{aligned}
\end{equation}
 The turn over state of Eq. (\ref{eq:n=five}) is
\begin{equation}
\label{eq:n=checkpsi52}
\begin{aligned}
\ket{\psi^5}^{To} & =\ket{\psi^3}^{To}\frac{\ket{00}+\ket{11}}{2}+\ket{\psi'^3}^{To}\frac{\ket{01}-\ket{10}}{2}\\
 & =\frac{1}{4\sqrt{2}}\bigg\{\Big[\big(\ket{00}+\ket{11}\big)\big(\ket{0}+\ket{1}\big)+\big(\ket{10}-\ket{01}\big)\big(\ket{0}-\ket{1}\big)\Big]\big(\ket{00}+\ket{11}\big)\\
 & \quad +\Big[\big(\ket{00}+\ket{11}\big)\big(\ket{0}-\ket{1}\big)-\big(\ket{10}-\ket{01}\big)\big(\ket{0}+\ket{1}\big)\Big]\big(\ket{01}-\ket{10}\big)\bigg\}\\
 & =\frac{1}{4\sqrt{2}}\bigg\{\big(\ket{00}+\ket{11}\big)\Big[\big(\ket{0}+\ket{1}\big)\big(\ket{00}+\ket{11}\big)+\big(\ket{0}-\ket{1}\big)\big(\ket{01}-\ket{10}\big)\Big]\\
 & \quad +\big(\ket{10}-\ket{01}\big)\Big[\big(\ket{0}-\ket{1}\big)\big(\ket{00}+\ket{11}\big)-\big(\ket{0}+\ket{1}\big)\big(\ket{01}-\ket{10}\big)\Big]\bigg\}\\
 & =\frac{\ket{00}+\ket{11}}{2}\ket{\psi^3}^{To}+\frac{\ket{10}-\ket{01}}{2}\ket{\psi'^3}^{To}=\frac{\ket{00}+\ket{11}}{2}\ket{\psi^3}+\frac{\ket{10}-\ket{01}}{2}\ket{\psi'^3}=\ket{\psi^5}.
\end{aligned}
\end{equation}
We can show $\ket{\psi'^5}^{To}=\ket{\psi'^5}$ in the same way.

With mathematical induction, we can deduce $\ket{\psi^7}$ and $\ket{\psi'^7}$ satisfy the same statements, and so do $(2N+1)$-qubit. So the $(2N+1)$-qubit final states we constructed in Sec. \ref{sec:sub3} obey Theorem \ref{th:L=3}. We should notice that, the back-up basis $\ket{\psi'^{2N+1}}$ does not obey Theorem \ref{th:L=3}, so it can not be generated in the symmetric setup.

~\\
\indent\textbf{2.The $2N$-qubit case}

Now we consider the two-qubit basis in Eq. (\ref{eq:n=bases2}),
\begin{equation}
\label{eq:n=checkpsi2}
\begin{aligned}
\widetilde{\ket{k_+}} & =\frac{1}{2}\Big[\big(\ket{11}+\ket{00}\big)+i(\ket{10}+\ket{01}\big)\Big]=\ket{k_+},\\
\ket{k_+}^{To} & =\frac{1}{2}\Big[\big(\ket{00}+\ket{11}\big)+i(\ket{10}+\ket{01}\big)\Big]=\ket{k_+},\\
\widetilde{\ket{k_-}} & =\frac{1}{2}\Big[i\big(\ket{11}+\ket{00}\big)+(\ket{10}+\ket{01}\big)\Big]=\ket{k_-},\\
 \ket{k_-}^{To} & =\frac{1}{2}\Big[i\big(\ket{00}+\ket{11}\big)+(\ket{10}+\ket{01}\big)\Big]=\ket{k_-}.
\end{aligned}
\end{equation}
The four-qubit basis in Eq. (\ref{eq:n=4qubits}) and Eq. (\ref{eq:n=4bases}) satisfy,
\begin{equation}
\label{eq:n=checkpsi4}
\begin{aligned}
 & \widetilde{\ket{\psi^4}}=\frac{\ket{11}+\ket{00}}{2}\widetilde{\ket{k_+}}+\frac{\ket{10}+\ket{01}}{2}\widetilde{\ket{k_-}}=\ket{\psi^4},\\
 & \widetilde{\ket{\psi'^4}}=\frac{\ket{11}+\ket{00}}{2}\widetilde{\ket{k_-}}+\frac{\ket{10}+\ket{01}}{2}\widetilde{\ket{k_+}}=\ket{\psi'^4}.
\end{aligned}
\end{equation}
\begin{equation}
\label{eq:n=checkpsi42}
\begin{aligned}
\ket{\psi^4}^{To} & =\ket{k_+}^{To}\frac{\ket{00}+\ket{11}}{2}+\ket{k_-}^{To}\frac{\ket{10}+\ket{01}}{2}\\
 & =\frac{1}{4}\bigg\{\Big[\big(\ket{00}+\ket{11}\big)+i\big(\ket{01}+\ket{10}\big)\Big]\big(\ket{00}+\ket{11}\big)\\
 & +\Big[i\big(\ket{00}+\ket{11}\big)+\big(\ket{01}+\ket{10}\big)\Big]\big(\ket{10}+\ket{01}\big)\bigg\}\\
 & =\frac{1}{4}\bigg\{\big(\ket{00}+\ket{11}\big)\Big[\big(\ket{00}+\ket{11}\big)+i\big(\ket{10}+\ket{01}\big)\Big]\\
 & +\big(\ket{01}+\ket{10}\big)\Big[i\big(\ket{00}+\ket{11}\big)+\big(\ket{10}+\ket{01}\big)\Big]\bigg\}\\
 & =\frac{\ket{00}+\ket{11}}{2}\ket{k_+}+\frac{\ket{01}+\ket{10}}{2}\ket{k_-}=\ket{\psi^4}.
\end{aligned}
\end{equation}
We can show $\ket{\psi'^4}^{To}=\ket{\psi'^4}$ similarly.

Inducing in the same way,
$2N$-qubit basis also satisfy Theorem \ref{th:L=3}. So $2N$-qubit final states constructed in Sec. \ref{sec:sub4} can be produced in the symmetric setup. In this case, all back-up bases $\ket{\psi'^{2N}}$ obey Theorem \ref{th:L=3} too.

In this section, we have extended the gravitational entanglement to multiqubit. If we choose appropriate parameters in the apparatus, the masses can be transformed to GHZ-type entangled states.
Since the number of independent phases are constrained in symmetric setup, the apparatus can not produce W-type states. W-type states need more freedom of phases, that means less symmetry in the apparatus.
 For example, if we change the distances $d$ between masses in Fig. \ref{fig:1}(i), such as $d_1$ and $d_2$, the phases of each item are independent,
 \begin{equation}
\label{eq:n=three14}
 \begin{aligned}
 \ket{\psi_6}= & \frac{1}{2\sqrt{2}}\big(e^{i\phi'_1}\ket{000}+e^{i\phi'_2}\ket{001}+e^{i\phi'_3}\ket{010}+e^{i\phi'_4}\ket{100}\\
  & +e^{i\phi'_5}\ket{110}
+e^{i\phi'_6}\ket{101}+e^{i\phi'_7}\ket{011}+e^{i\phi'_8}\ket{111}\big).
\end{aligned}
\end{equation}
If we choose the phases as $e^{i\phi'_1}=e^{i\phi'_2}=e^{i\phi'_3}=e^{i\phi'_4}=e^{i\phi'_8}=1$, $e^{i\phi'_5}=-i$, $e^{i\phi'_6}=-1$, $e^{i\phi'_7}=i$, the modified apparatus will produce W-type states,
\begin{equation}
\label{eq:n=three11}
\begin{aligned}
\ket{\psi_6}= & \frac{1}{2\sqrt{2}}\big(\ket{000}+\ket{001}+\ket{010}+\ket{100}-i\ket{110}
-\ket{101}+i\ket{011}+\ket{111}\big)\\
 = & \frac{1}{2\sqrt{2}}\Big[\big(\ket{0}+\ket{1}\big)_{A_1}\otimes\ket{00}_{A_2A_3}+\big(\ket{0}-\ket{1}\big)_{A_1}\otimes\ket{01}_{A_2A_3}\\
  &  +\big(\ket{0}-i\ket{1}\big)_{A_1}\otimes\ket{1}_{A_2}\otimes\big(\ket{0}+i\ket{1}\big)_{A_3}\Big].
\end{aligned}
\end{equation}
The state in $\ket{\psi_6}$ can be transformed to W states under SLOCC.

\section{Measure Of Entanglement}
\label{sec:gm}
In this section, we analyse and measure the gravitational entanglement by the GM and negativity of entanglement.
GM is the closest distance between an entangled state and the set of separable states \cite{16,17},
\begin{equation}
\label{eq:n=gm}
\begin{split}
 \Lambda^2(\rho)=\mathop{max}\limits_{\ket{\varphi}\in PRO}\bra{\varphi}\rho\ket{\varphi},\\
 G(\rho)=-2\log\Lambda(\rho).
\end{split}
\end{equation}
Here the logarithm has base two.

We just derive GM of tripartite system for simplicity.
Suppose the three-qubit product states be
\begin{equation}
\label{eq:n=tris}
\begin{split}
 \ket{\phi}=(\cos\alpha\ket{0}+e^{i\theta}\sin\alpha\ket{1})_{A_1}\otimes(\cos\beta\ket{0}+e^{i\eta}\sin\beta\ket{1})_{A_2}\otimes
 (\cos\gamma\ket{0}+e^{i\omega}\sin\gamma\ket{1})_{A_3}.
\end{split}
\end{equation}
Since there are too many parameters in the product states, we will consider the symmetric form of the entangled state for simplicity in Sec. \ref{sec:sub4-1}, and study the general form in Sec. \ref{sec:sub4-2}. The conclusions are presented in Theorem. \ref{th:L=2}.
\begin{theorem}
\label{th:L=2}	
When we use GM to measure the entanglement, the gravity induced entanglement can produce stable and robust entanglement for tripartite system. The robust entanglement appears generally at $\Delta\varphi_3\in[\frac{11\pi}{16},\frac{21\pi}{16}]$ and the corresponding $G(\rho)$ is close to one.
\end{theorem}
We will show it in next subsections.

\subsection{Symmetric situation}
\label{sec:sub4-1}
First we consider the entangled states in Eq. (\ref{eq:n=three2}) be in symmetric form, that means $e^{i\Delta\varphi_2}=e^{i\Delta\varphi_3}$. The states can be written as
\begin{equation}
\label{eq:n=gm2}
\begin{split}
 \ket{\psi}=\frac{e^{i\varphi_1}}{2\sqrt{2}}\Big[\ket{000}+\ket{111}
   +e^{i\Delta\varphi_3}\big(\ket{001}+\ket{011}+
   \ket{100}+\ket{110}+\ket{010}+\ket{101}\big)\Big].
\end{split}
\end{equation}

As been showed in Proposition 4 of \cite{17}, the closest product state to any symmetric state is necessarily symmetric.
So the state in Eq. (\ref{eq:n=tris}) should be symmetric too (which means $\alpha=\beta=\gamma$, $\theta=\omega=\eta$), can be written as $\ket{\phi}=(\cos\alpha\ket{0}+e^{i\theta}\sin\alpha\ket{1})^{\otimes 3}$.

The inner product of the entangled state and the product state is
\begin{equation}
\label{eq:n=inner1}
\begin{split}
 \braket{\phi}{\psi}=\frac{e^{i\varphi_1}}{2\sqrt{2}}\bigg[\cos^3\alpha+\sin^3\alpha e^{-3i\theta}
 +3e^{i\Delta\varphi_3}\cos\alpha \sin\alpha e^{-i\theta}(\cos\alpha+\sin\alpha e^{-i\theta})\bigg].
\end{split}
\end{equation}
Its modular square is
\begin{equation}
\label{eq:n=abs2}
\begin{aligned}
  \lvert\braket{\phi}{\psi}\rvert^2=\frac{1}{8}\bigg\{ & \cos^6\alpha+\cos^5\alpha \sin\alpha \big[6\cos(\Delta\varphi_3-\theta)\big]+\cos^4\alpha \sin^2\alpha\big[6\cos(\Delta\varphi_3-2\theta)+9\big]\\
  & +\cos^3\alpha \sin^3\alpha(18\cos\theta+2\cos 3\theta)+\cos^2\alpha \sin^4\alpha\big[6\cos(\Delta\varphi_3+2\theta)+9\big]\\
 & +\cos\alpha \sin^5\alpha \big[6\cos(\Delta\varphi_3+\theta)\big]+\sin^6\alpha\bigg\}.\\
\end{aligned}
\end{equation}

\begin{figure}[h!]
 \begin{minipage}{0.33\linewidth}
  \centering
        \centerline{
        \includegraphics[width=1\textwidth]{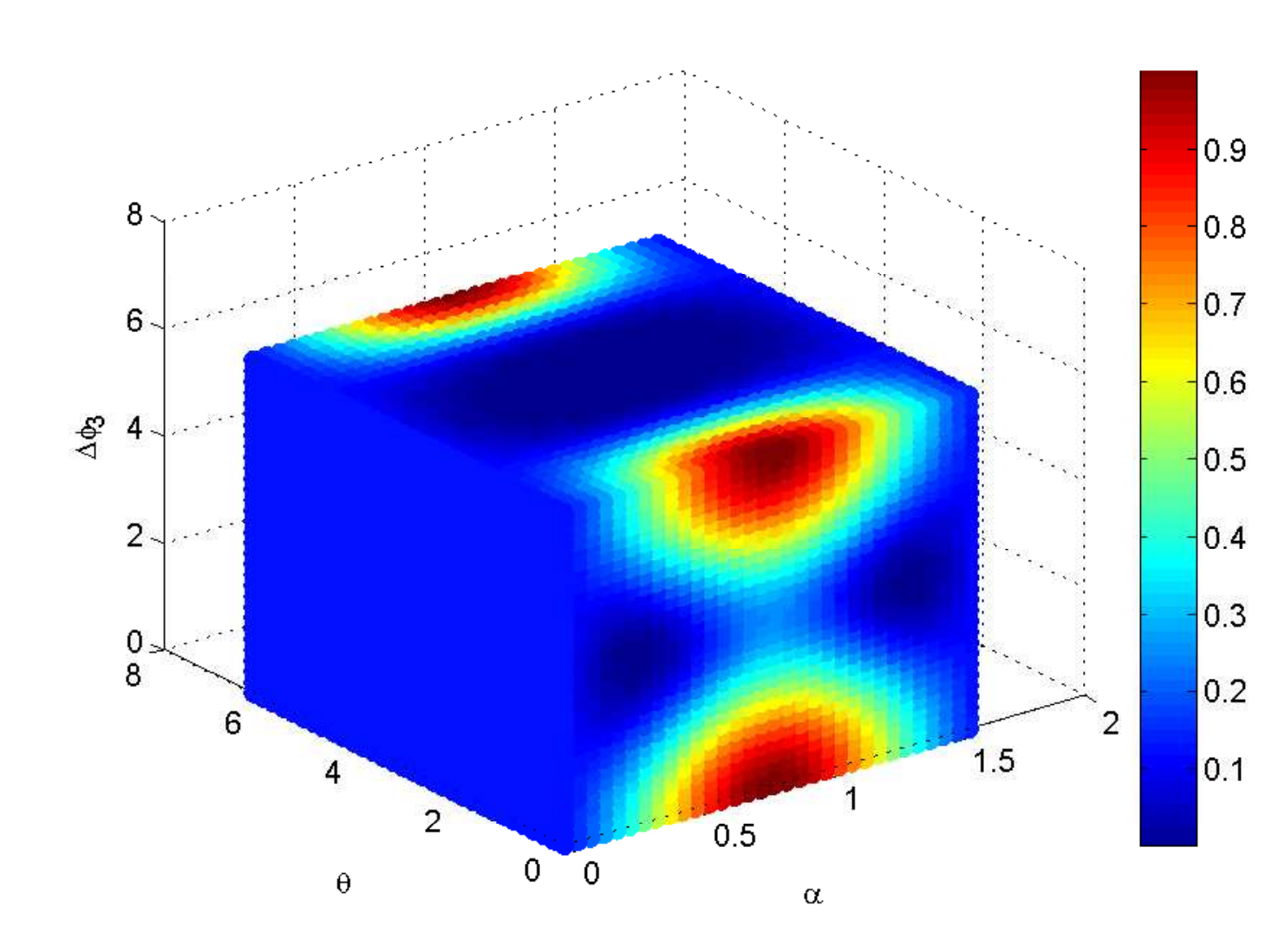}
        }
        \centerline{(i)}
    \end{minipage}
    \begin{minipage}{0.33\linewidth}
        \centering
        \centerline{
        \includegraphics[width=1\textwidth]{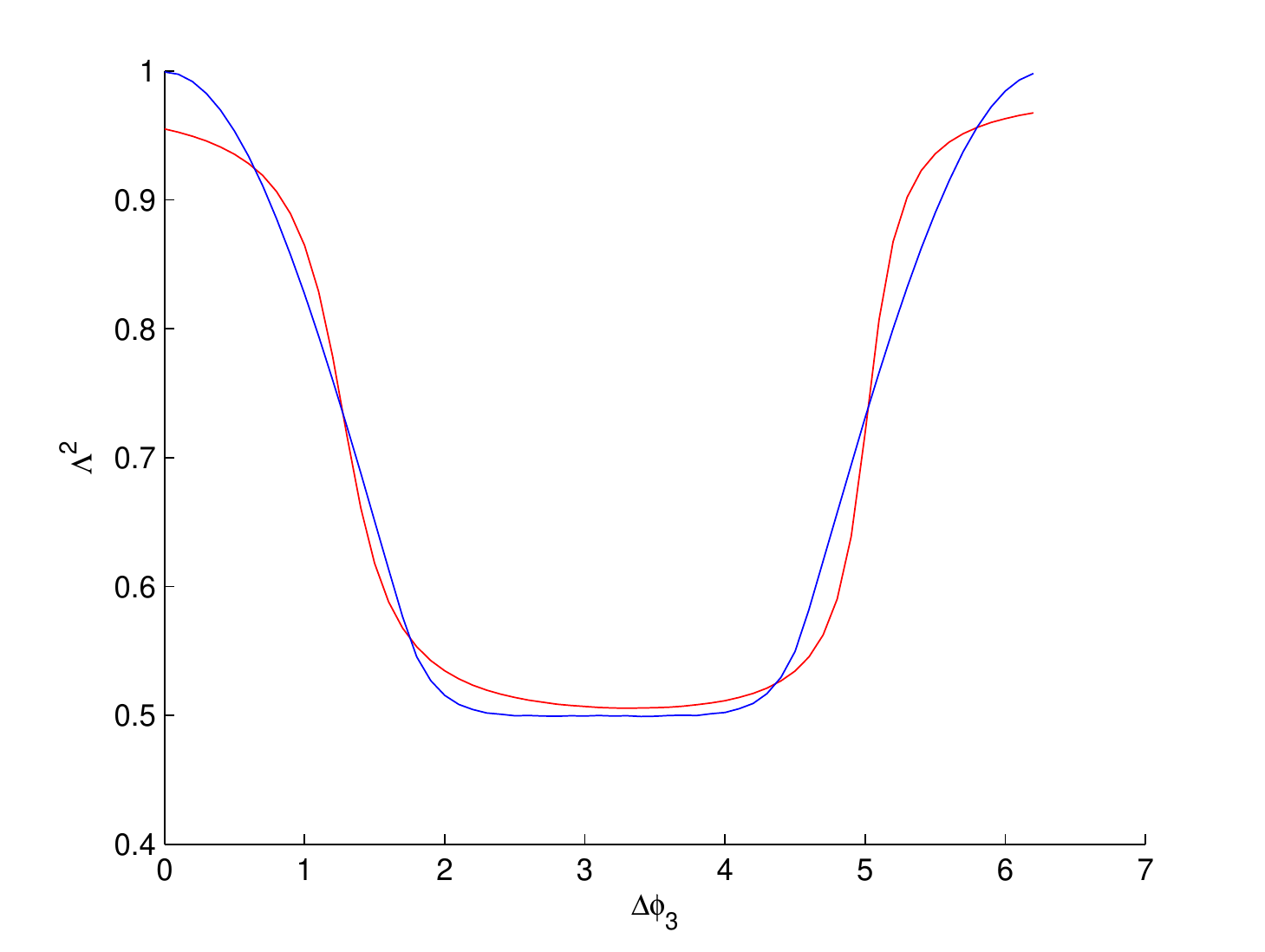}
        }
        \centerline{(ii)}
    \end{minipage}
\caption{(i) The colour stands for $\lvert\braket{\phi}{\psi}\rvert^2$, as the function of $\alpha$, $\theta$ and $\Delta\varphi_3$ in one period, the colour bar gives out approximate value of $\lvert\braket{\phi}{\psi}\rvert^2$.
(ii) The blue curve represents the $\Lambda^2$ as the function of $\Delta\varphi_3$, and the red curve represents the data fitting function in Eq. (\ref{eq:n=lamda2}).}
\label{fig:6}
\end{figure}
Eq. (\ref{eq:n=abs2}) is a function of $\alpha$, $\theta$ and $\Delta\varphi_3$, which is presented in Fig. \ref{fig:6} (i). As the colour bar shows, for $\Delta\varphi_3$ close to zero, $\Lambda^2$ get close to one, and the nearest product state appears around $\alpha=\pi/4$, $\theta=0$. The criterion proposed in Sec. \ref{sec:sub1} shows that the state in Eq. (\ref{eq:n=gm2}) becomes a separable state for $\Delta\varphi_3=0$, so GM is vanishing and $\Lambda^2=1$ ($G(\rho)$=0). At the beginning of the gravitational interaction, the relative evolution phase $\Delta\varphi_3$ is small, and the entangled state in Eq. (\ref{eq:n=gm2}) is near the product state, so $\Lambda^2$ is big. With the interaction going on, the entanglement of Eq. (\ref{eq:n=gm2}) increases, the distance between entangled state and the nearest product state increases, and $\Lambda^2$ falls down.
\begin{table}
\caption{\label{tab:table1} The data of $\Lambda^2$ and the parameters of the nearest product states for some certain relative evolution phases $\Delta\varphi_3$.}
\begin{ruledtabular}
\begin{tabular}{cccccccccccccccc}
$\Delta\varphi_2$ & $\frac{\pi}{8}$ & $\frac{\pi}{4}$ & $\frac{3\pi}{8}$ & $\frac{\pi}{2}$ & $\frac{5\pi}{8}$ & $\frac{3\pi}{4}$ & $\frac{7\pi}{8}$ & $\pi$  & $\frac{9\pi}{8}$  & $\frac{5\pi}{4}$ & $\frac{11\pi}{8}$ & $\frac{3\pi}{2}$ & $\frac{13\pi}{8}$ & $\frac{7\pi}{4}$ & $\frac{15\pi}{8}$\\
$\Lambda^2$ & 0.97 & 0.89 & 0.77 & 0.625 & 0.52 & 0.50 & 0.50 & 0.50  & 0.50 & 0.50 & 0.52 & 0.625 & 0.77 & 0.89 & 0.97\\
$\alpha$ & $\frac{\pi}{4}$ & $\frac{\pi}{4}$ & $\frac{\pi}{4}$ & $\frac{\pi}{4}$ & $\frac{\pi}{32}$ & $\frac{\pi}{32}$ & $\frac{7\pi}{32}$ & $\frac{\pi}{4}$ & $\frac{7\pi}{32}$ & $\frac{5\pi}{16}$ & $\frac{5\pi}{16}$ & $\frac{\pi}{4}$ & $\frac{\pi}{4}$ & $\frac{\pi}{4}$ & $\frac{\pi}{4}$\\
$\theta$ & 0 & 0 & 0 & 0 & $\frac{27\pi}{16}$ & $\frac{25\pi}{16}$ & $\frac{\pi}{2}$ & $\frac{\pi}{2}$ & $\frac{3\pi}{2}$ & $\frac{7\pi}{16}$ & $\frac{5\pi}{16}$ & 0 & 0 & 0 & 0\\
\end{tabular}
\end{ruledtabular}
\end{table}

 With numerical computation, the blue curve in Fig. \ref{fig:6} (ii) describes $\Lambda^2$ as the function of $\Delta\varphi_3$ only, the nearest product states for each $\Delta\varphi_3$ are different, the corresponding $\alpha$ and $\theta$ were partly listed in Table \ref{tab:table1}.
 Both ends of the curve ($\Delta\varphi_3<\frac{\pi}{4}$ or $\Delta\varphi_3>\frac{7\pi}{4}$) correlate to weak entanglement with $\Lambda^2$ close to one. The nearest product states have stable phases $\alpha=\frac{\pi}{4}$ and $\theta=0$. In the rapid change ranges ($\Delta\varphi_3\in [\frac{\pi}{4}, \frac{\pi}{2}]$ and $[\frac{3\pi}{2}, \frac{7\pi}{4}]$) of the curve, $\Lambda^2$ changes rapidly. The nearest product states also are $\ket{\phi}=(\frac{\ket{0}+\ket{1}}{\sqrt{2}})^{\otimes 3}$.
 The middle range of the curve ($\Delta\varphi_3\in (\frac{\pi}{2}, \frac{3\pi}{2})$) represents the final states entangled strongly, with small $\Lambda^2$ around $0.5$. However, the nearest product states are oscillating. In this range, entanglement is great and stable, sensitive to witnesses. It is an important region for studying the multipartite entanglement. GM in Eq. (\ref{eq:n=gm}) becomes $G(\rho)=1$ in this region. Nevertheless, the apparatus can only generate small relative evolution phase $\Delta\varphi_3$ in \cite{1}, since the interaction can not last for a long time because of decoherence. If we want to detect robust entangled state, we can consider heavier masses as in \cite{2}. When we detect some entangled states with certain phases $\Delta\varphi_3$, corresponding $\alpha$ and $\theta$ are listed in Table \ref{tab:table1}, or we can use the fitting function of the curve to compute.
\begin{equation}
\label{eq:n=lamda2}
\begin{split}
  \Lambda^2=0.164[\arctan(5.71\Delta\varphi_2-28.68)+\arctan(-3.79\Delta\varphi_2+4.83)]+0.98.\\
\end{split}
\end{equation}
The red curve in Fig. \ref{fig:6} (ii) presents the theoretical result.

\subsection{General Situation}
\label{sec:sub4-2}
In this subsection, we consider the product states be chosen as Eq. (\ref{eq:n=tris}), the inner product becomes
\begin{equation}
\label{eq:n=inner}
\begin{aligned}
 \braket{\phi}{\psi}= & \frac{e^{i\varphi_1}}{2\sqrt{2}}\Big[\cos\alpha \cos \beta \cos\gamma+\sin\alpha \sin\beta \sin\gamma e^{-i(\theta+\eta+\omega)}\\
& +e^{i\Delta\varphi_2}\big(\cos\alpha \cos\beta \sin\gamma e^{-i\omega}+\cos\alpha \sin\beta \sin\gamma e^{-i(\eta+\omega)}+\sin\alpha \cos\beta \cos\gamma e^{-i\theta}\\
& +\sin\alpha \sin\beta \cos\gamma e^{-i(\theta+\eta)}\big)+e^{i\Delta\varphi_3}\big(\cos\alpha \sin\beta \cos\gamma e^{-i\eta}+\sin\alpha \cos\beta \sin\gamma e^{-i(\theta+\omega)}\big)\Big].
\end{aligned}
\end{equation}

\begin{figure}[h!]
 \begin{minipage}{0.33\linewidth}
  \centering
        \centerline{
        \includegraphics[width=1\textwidth]{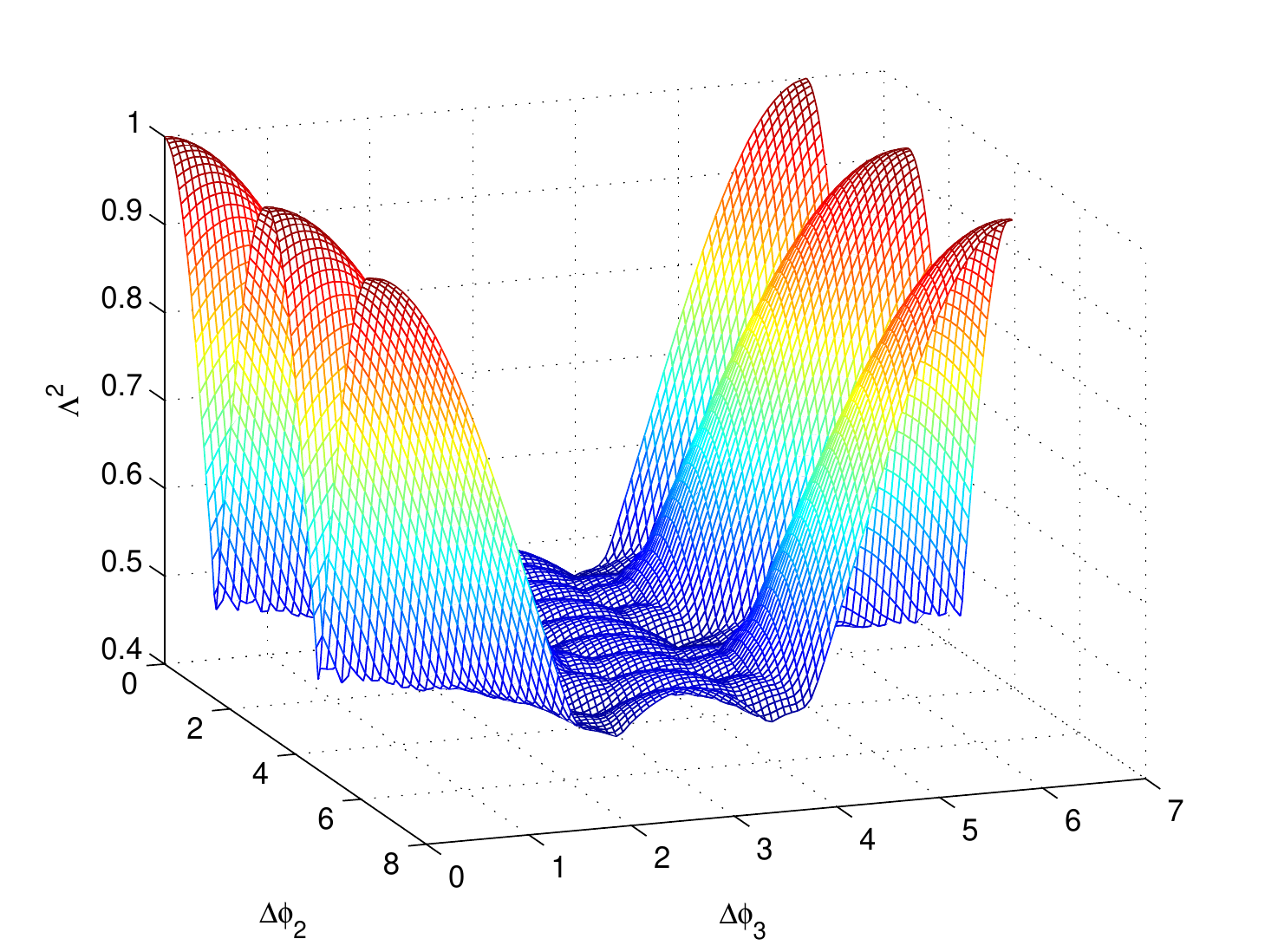}
        }
        \centerline{(i)}
    \end{minipage}
    \begin{minipage}{0.33\linewidth}
        \centering
        \centerline{
        \includegraphics[width=1\textwidth]{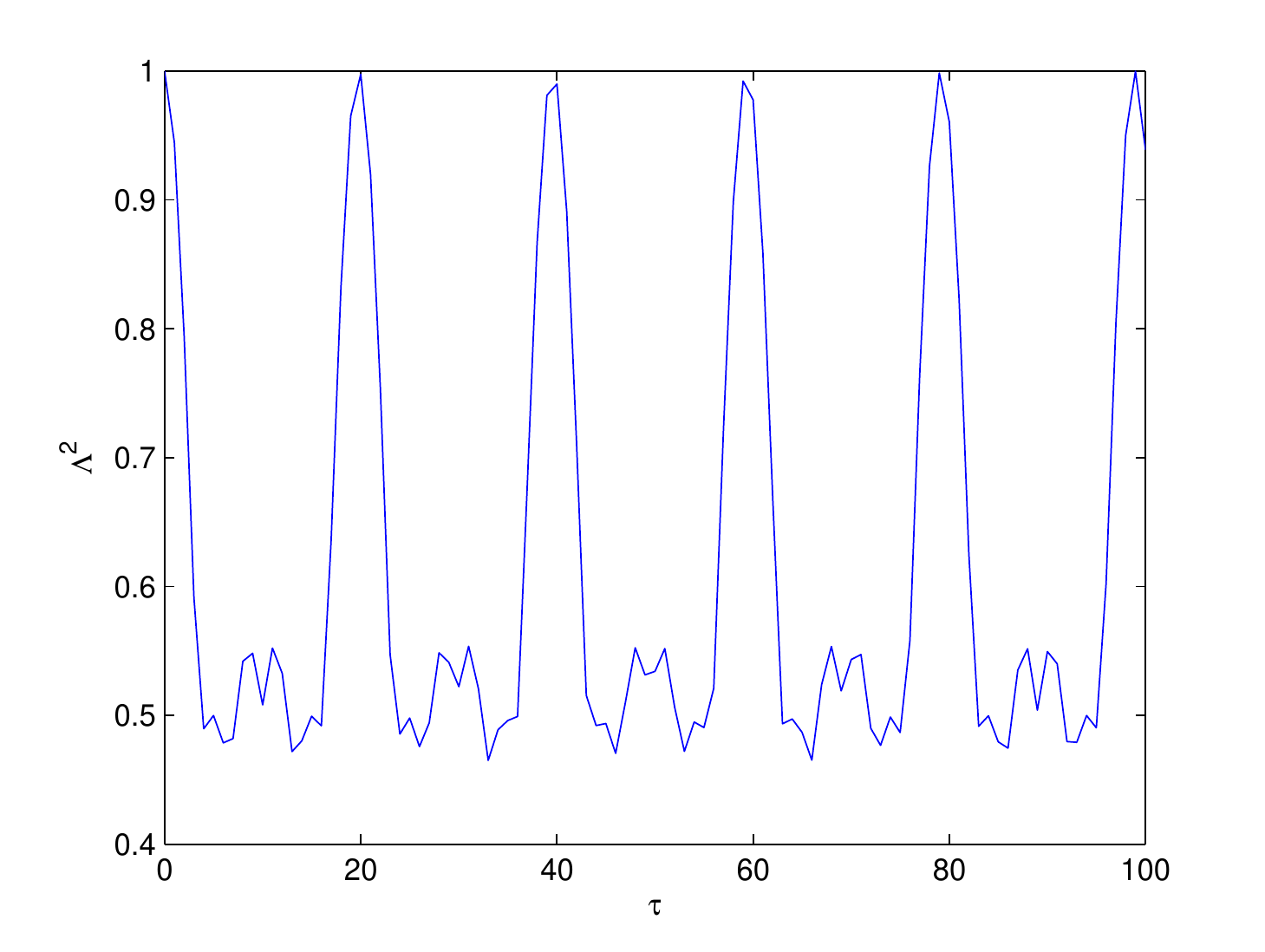}
        }
        \centerline{(ii)}
    \end{minipage}
\caption{(i)The maximum of modular square $\Lambda^2$ as the function of $\Delta\varphi_2$ and $\Delta\varphi_3$ in one period. (ii)$\Lambda^2$ as the function of free falling time $\tau$.}
\label{fig:7}
\end{figure}

Since there are many parameters in Eq. (\ref{eq:n=inner}), we study it by numerical analysis. We just present how $\Lambda^2$ trends with $\Delta\varphi_2$ and $\Delta\varphi_3$ in Fig. \ref{fig:7}, because it is hard to give the exact function form.
The states in Eq. (\ref{eq:n=three2}) become product states when $e^{i\Delta\varphi_2}=\pm 1$ and $e^{i\Delta\varphi_3}=1$, GM vanishes and the entanglement is weak in adjacent regions (the red parts in Fig. \ref{fig:7}). The nearest product states are $\ket{\phi}=\frac{1}{2\sqrt{2}}(\ket{0}\pm \ket{1})_{A_1}\otimes(\ket{0}+\ket{1})_{A_2}\otimes(\ket{0}\pm \ket{1})_{A_3}$, with $\alpha=\beta=\gamma=\pi/4$ and $\theta=\omega=\eta=0$ (or $\theta=\eta=\pi$, $\omega=0$).
There are also some regions presenting robust and stable entanglement in Fig. \ref{fig:7} (the blue parts). For different $\Delta\varphi_2$, the ranges of $\Delta\varphi_3$ (corresponding to blue parts) are a little different, we choose the intersection $\Delta\varphi_3\in[\frac{11\pi}{16},\frac{21\pi}{16}]$, it is valid for arbitrary $\Delta\varphi_2$. In these regions, $\Lambda^2$ is a little smaller than $0.5$, the maximum $G(\rho)$ is about $1.14$. The corresponding nearest product states vary with relative evolution phases $\Delta\varphi_2$ and $\Delta\varphi_3$.
For example, when $\Delta\varphi_2=2\pi$ and $\Delta\varphi_3=\pi$, the nearest product states is $\ket{\phi}=\frac{1}{2\sqrt{2}}(\ket{0}+i \ket{1})_{A_1}\otimes(\ket{0}-i\ket{1})_{A_2}\otimes(\ket{0}+i\ket{1})_{A_3}$, with $\alpha=\beta=\gamma=\pi/4$ and $\theta=\eta=\pi/2$, $\omega=3\pi/2$.

If the parameters in the experiment are chosen as \cite{5}, $m_1, m_2, m_3\sim 10^{-14}kg$, $d\sim 200\mu m$ and $l>>d$.
We also consider $\Lambda^2$ as the function of free falling time $\tau$ in Fig. \ref{fig:7}(ii), the period is about 20 seconds.

\subsection{Negativity}
In this part, we obtain tripartite negativity \cite{27} to measure the entanglement in Eq. (\ref{eq:n=three2}). It is defined as
\begin{equation}
\label{eq:n=negativity}
\begin{aligned}
 N_{ABC}(\rho)=(N_{A-BC}N_{B-AC}N_{C-AB})^{\frac{1}{3}}.
\end{aligned}
\end{equation}
Here $N_{I-JK}=-2\sum \sigma_i(\rho^{TI})$ is the bipartite negativity,
$\sigma_i(\rho^{TI})$ is the negative eigenvalue of $\rho^{TI}$, the partial transpose of $\rho$ with respect to subsystem I.

\begin{figure}[h!]
 \begin{minipage}{0.33\linewidth}
  \centering
        \centerline{
        \includegraphics[width=1\textwidth]{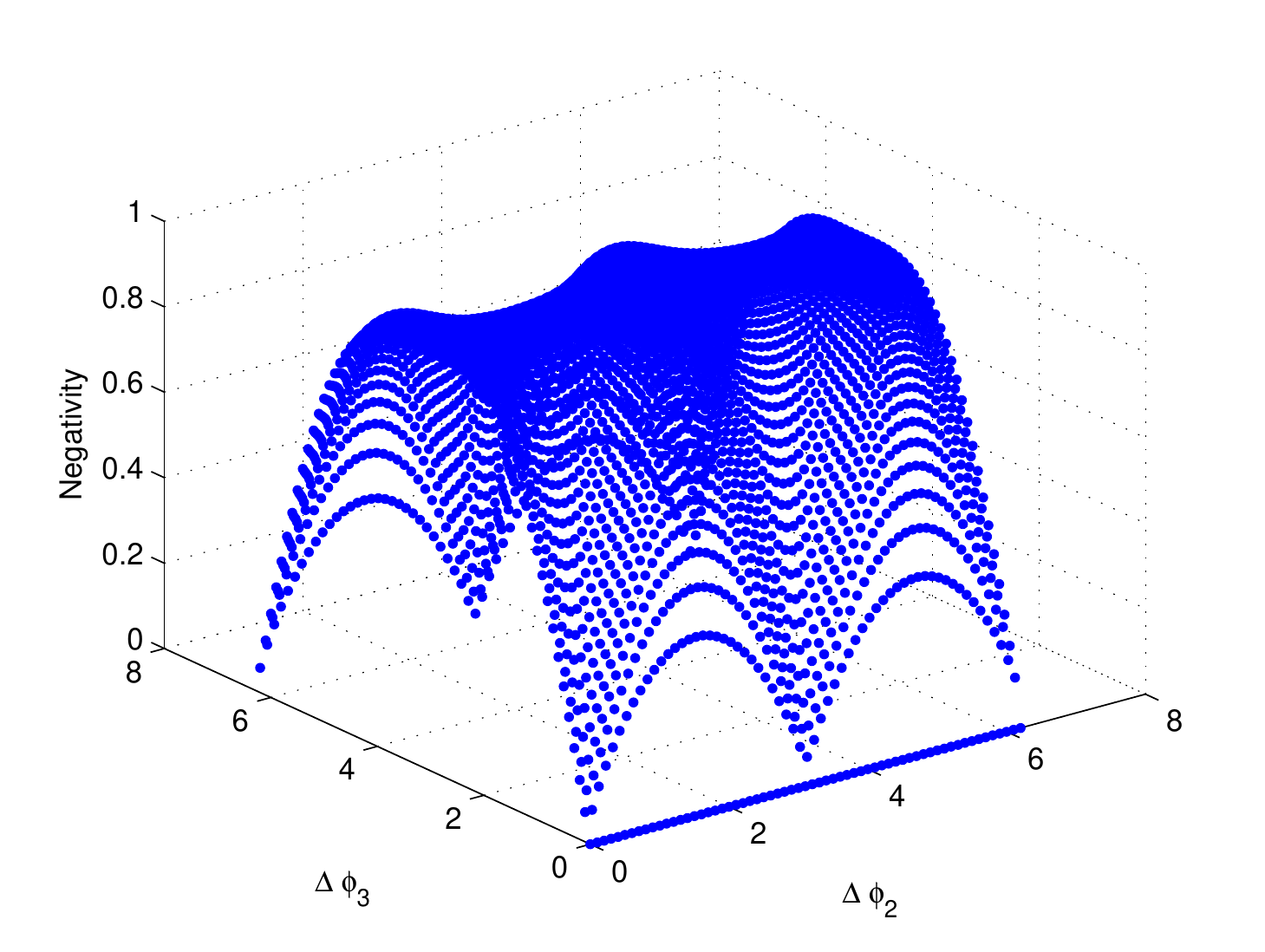}
        }
        \centerline{(i)}
    \end{minipage}
    \begin{minipage}{0.33\linewidth}
        \centering
        \centerline{
        \includegraphics[width=1\textwidth]{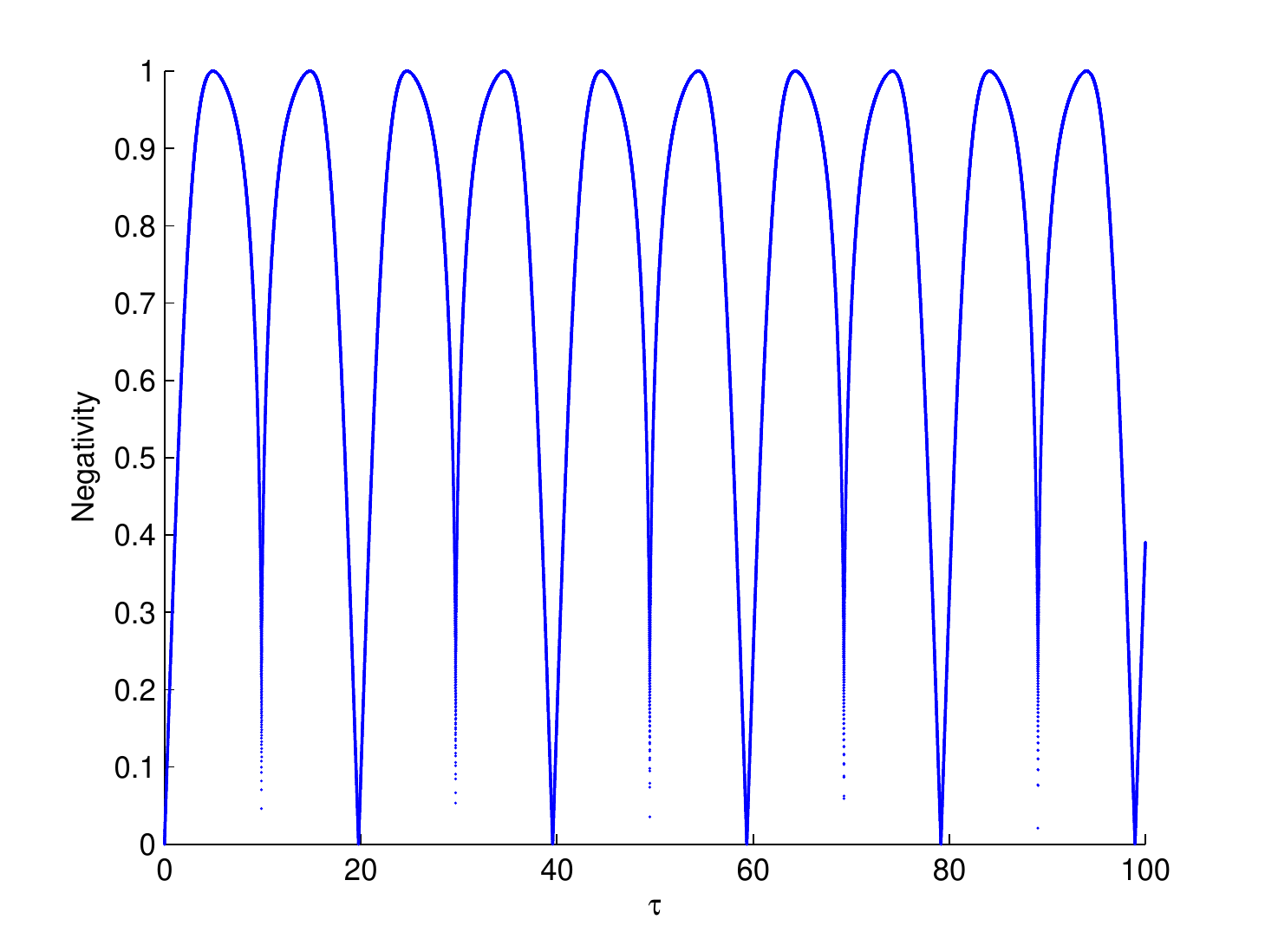}
        }
        \centerline{(ii)}
    \end{minipage}
\caption{(i)The tripartite negativity as the function of $\Delta\varphi_2$ and $\Delta\varphi_3$ in one period. (ii)The tripartite negativity as the function of free falling time $\tau$.}
\label{fig:8}
\end{figure}

The negativity is related to the evolution phases, as Fig. \ref{fig:8}(i) shows, $N=1$ corresponds to maximum entangled states with $\Delta\varphi_3=\pi$, they are GHZ states mentioned in Sec. \ref{sec:mq}.
Consider the experiment with the same parameters have mentioned above,
we present $N$ as the function of free falling time $\tau$ in Fig. \ref{fig:8}(ii). Negativity oscillates with $\tau$ for the same period as $\Lambda^2$. If we detect this change trend in experiment, it is a signal of quantum gravity.

\begin{figure}[h!]
 \begin{minipage}{0.33\linewidth}
  \centering
        \centerline{
        \includegraphics[width=1\textwidth]{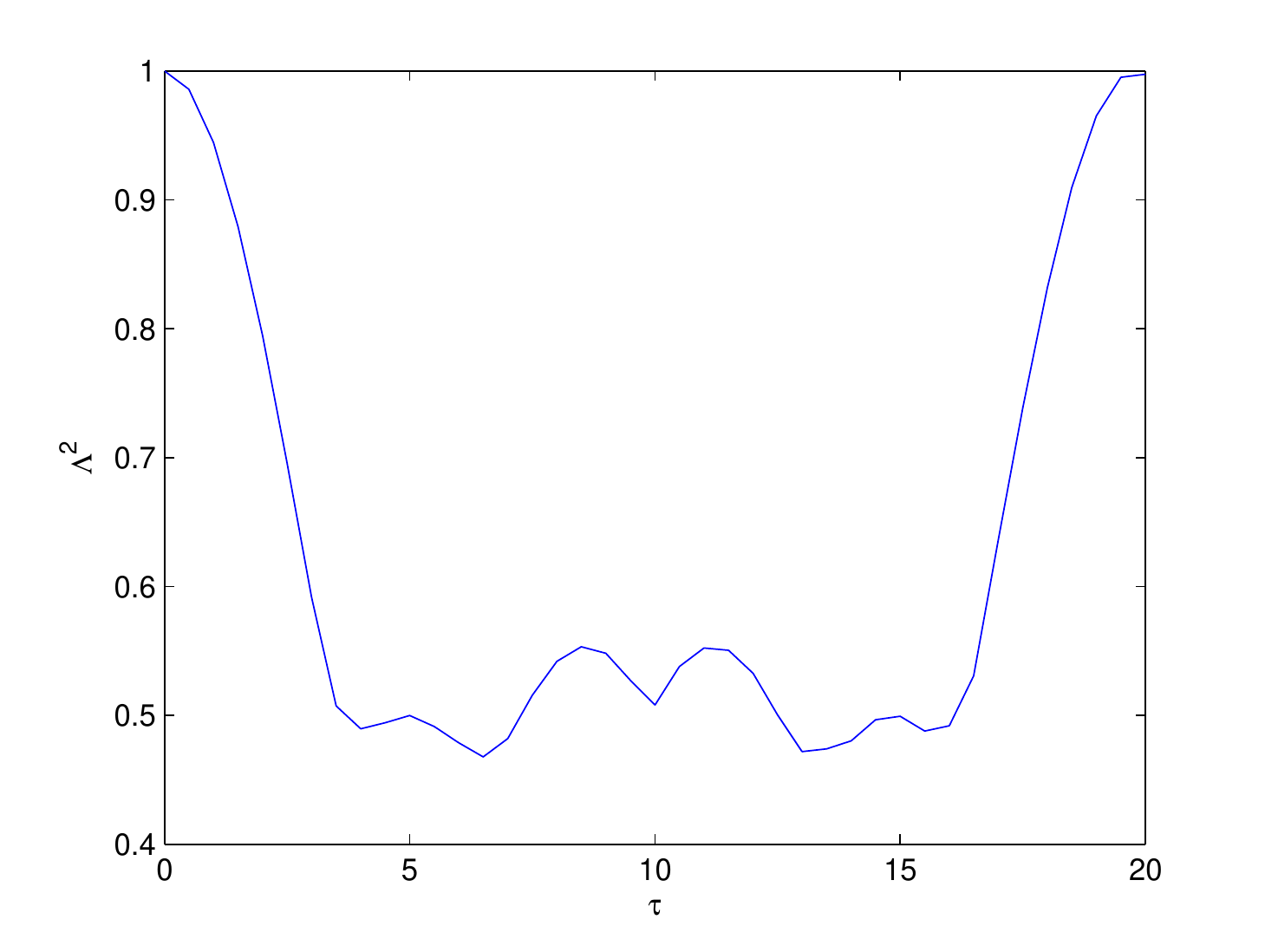}
        }
        \centerline{(i)}
    \end{minipage}
    \begin{minipage}{0.33\linewidth}
        \centering
        \centerline{
        \includegraphics[width=1\textwidth]{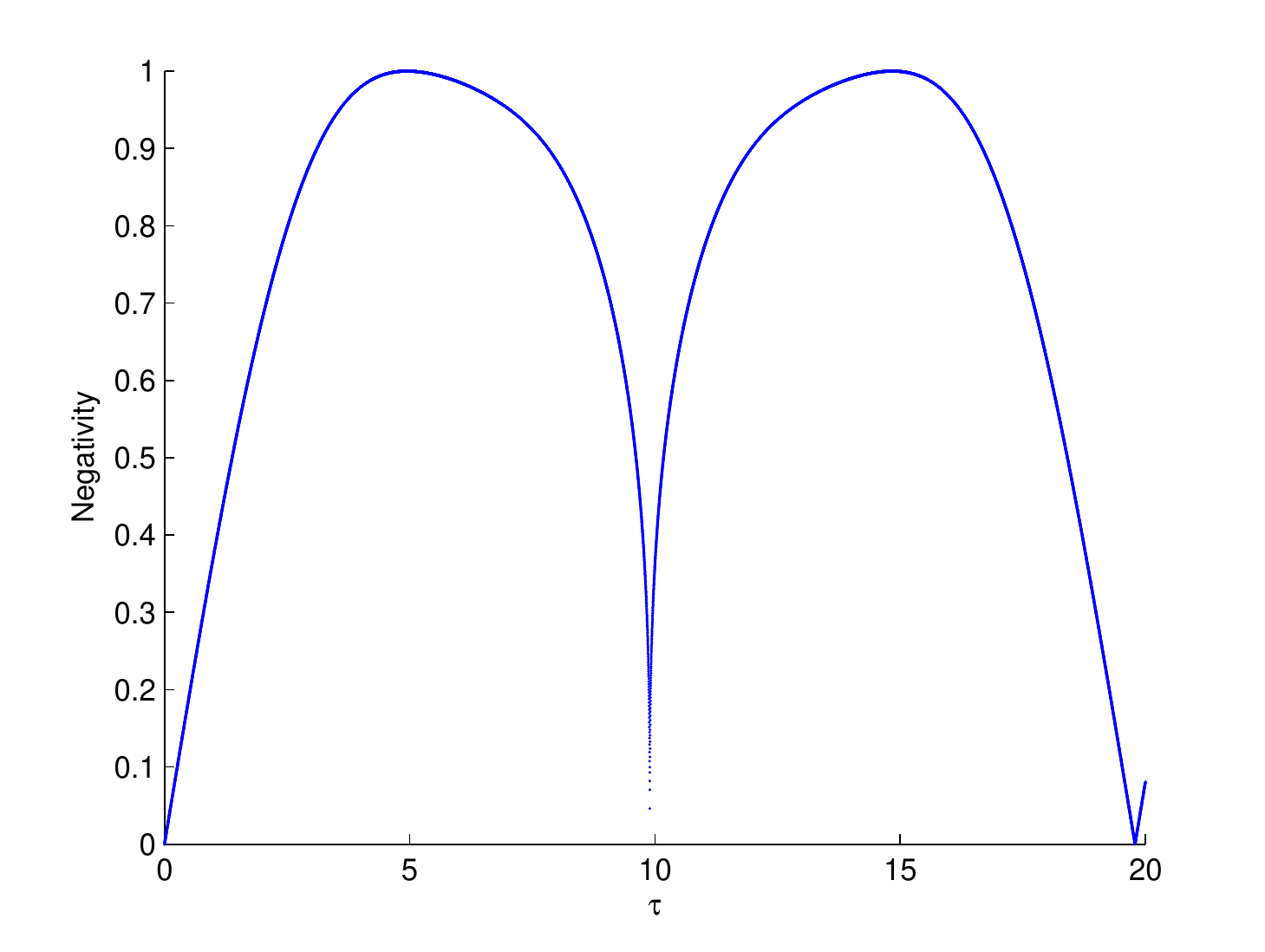}
        }
        \centerline{(ii)}
    \end{minipage}
\caption{(i)$\Lambda^2$ as the function of free falling time $\tau$ in one period. (ii)The tripartite negativity as the function of free falling time $\tau$ in one period.}
\label{fig:9}
\end{figure}

In Fig. \ref{fig:9}, we compare Fig. \ref{fig:7}(ii) and Fig. \ref{fig:8}(ii) in one period. With both GM and negativity measurement of entanglement, the gravity interaction generates strong entanglement when free falling time $\tau\in[5, 15]$ seconds. But when $\tau$ is around 10s, the negativity falls down rapidly, so if we want to inspect quantum gravity, this free falling time should be avoided.

In this section, we derived GM and negativity of the gravitational entanglement. They depend on the relative evolution phases $\Delta\varphi_2$ and $\Delta\varphi_3$. With numerical computation, we constructed the phase map of GM, it marks the degree of entanglement of the final states for each couple of relative evolution phases. We found that the robust entanglement appears generally at $\Delta\varphi_3\in[\frac{11\pi}{16},\frac{21\pi}{16}]$. And the negativity gives consistent results.
When we have found GM and negativity as the function of free falling time, the robust entangled states appears at $\tau\in[5, 15]$ seconds. Actually, if we can keep the masses stable for more than 1s, entanglement will be detectable. That reveals the quantum characters of gravity.

\section{Conclusions}
\label{sec:con}
In this article, we studied the multiqubit entanglement caused by gravity of neutral masses. The mutual gravitation interaction could transform the separable states into GHZ-type states. It suggested a way to construct $N$-qubit GHZ-type states by getting Bell states and $(N-2)$-qubit GHZ-type states entangled. This approach is not only feasible in gravity induced entanglement, but also can be applied in ions trap, because Coulomb force between ions acts in the same way as gravity except for a minus sign.
Since electromagnetic interaction is much stronger than gravity, the ions' entanglement is more feasible in experiment. Ions in a Paul trap can form a linear string, the positions of them satisfy the spatial symmetry in this work. For ions' string, the evolution phases of superpositions are caused by oscillating electric field and mutual coulomb force between ions.
With appropriate parameters, electric field induces common phase, and relative phases are only related to mutual interaction. So the results in this article can be applied in ions trap to involve more ions in the entangled states. That may be a feasible path to realize robust multiqubit entanglement and will play a part in quantum computation.

We also derived GM and negativity of three-qubit gravity induced entangled states. They measure the degree of entanglement for a certain final state. We constructed functions to describe the relationship of GM, negativity and the relative evolution phases. The phase map of GM enables us to pick out the ranges with robust entanglement. On the other hand, the measurements oscillate with free falling time, the period is valuable for entanglement detect. We will pay attention on these regions in the future study of multipartite entanglement. It is helpful in experiment design.

\section*{Data Availability}
All data supporting the findings of this study are available within the article or from the corresponding author upon reasonable request.

\section*{Acknowledgements}
SML thanks the interesting discussion with Professor Qilin Zhang and Mingguo Sun. MFL and LC were supported by the NNSF of China (Grant No. 11871089), and the Fundamental Research Funds for the Central Universities(Grant No. ZG216S2005).

\section*{Author Contributions}
Lin Chen and Shaomin Liu designed the methodology and developed the theoretical aspects. Shaomin Liu and Mengfan Liang performed calculation and wrote the paper. All authors discussed the results and contributed to refining the paper.

\vspace{12 pt}

\textbf{Competing Interests:} The authors declare no Competing Financial or Non-Financial Interests.

\bibliographystyle{unsrt}

\begin{thebibliography}{1}
\bibitem{18} Overstreet C, Asenbaum P, Curti J, Kim M, Kasevich MA. Observation of a gravitational Aharonov-Bohm effect. Science. 2022 Jan 14;375(6577):226-229. doi: 10.1126/science.abl7152. Epub 2022 Jan 13. PMID: 35025635.
\bibitem{19} Hohensee MA, Estey B, Hamilton P, Zeilinger A, M{\"u}ller H. Force-free gravitational redshift: proposed gravitational Aharonov-Bohm experiment. Phys Rev Lett. 2012 Jun 8;108(23):230404. doi: 10.1103/PhysRevLett.108.230404. Epub 2012 Jun 7. PMID: 23003927.
\bibitem{1} S. Bose, A. Mazumdar, G.W. Morley, H. Ulbricht, M. Toro\v{s}, M. Paternostro, A.A. Geraci, P.F. Barker, M.S. Kim, G. Milburn. Spin engtanglement witness for quantum gravity. Phys. Rev. Lett. 119, 240401 (2017).
\bibitem{2} C. Marletto, V. Vedral. Gravitationally-induced entanglement between two massive particles is sufficient evidence of quantum effects in gravity. Phys. Rev. Lett. 119, 240402 (2017).
\bibitem{11} Andr\'{e} Gro$\beta$ardt. Gravitational entanglement and the mass contribution of internal energy in nonrelativistic quantum systems. arXiv: 2204.03322v1 (2022).
\bibitem{15} M. Christodoulou, C. Rovelli. On the possiblity of laboratory evidence for quantum superposition of geometries. Phys. Lett. B 792, 64 (2018).
\bibitem{10} C Marletto et al. Quantum-gravity effects could in principle be witnessed in neutrino-like oscillations. New J. Phys. 20 083011 (2018).
\bibitem{5} Nguyen H C, Bernards F. Entanglement dynamics of two mesoscopic objects with gravitational interaction[J]. The European Physical Journal D, 2020, 74(4).
\bibitem{29} Barredo D, Lienhard V, SD L\'{e}s\'{e}leuc, et al. Synthetic three-dimensional atomic structures assembled atom by atom. 10.1038/S41586-018-0450-2[P]. 2017.
\bibitem{28} Barnes, K., Battaglino, P., Bloom, B.J. et al. Assembly and coherent control of a register of nuclear spin qubits. Nat Commun 13, 2779 (2022).
\bibitem{20} Thomas, P., Ruscio, L., Morin, O. et al. Efficient generation of entangled multiphoton graph states from a single atom. Nature 608, 677-681 (2022).
\bibitem{21} Sackett, C. A. et al. Experimental entanglement of four particles. Nature 404, 256-259 (2000).
\bibitem{22} Gong, M. et al. Genuine 12-qubit entanglement on a superconducting quantum processor. Phys. Rev. Lett. 122, 110501 (2019).
\bibitem{23} Besse, JC., Reuer, K., Collodo, M.C. et al. Realizing a deterministic source of multipartite-entangled photonic qubits. Nat. Commun. 11, 4877 (2020).
\bibitem{24} Pogorelov, I. et al. Compact ion-trap quantum computing demonstrator. PRX Quantum 2, 020343 (2021).
\bibitem{30} Schut M, Tilly J, Marshman R J, et al. Improving resilience of the Quantum Gravity Induced Entanglement of Masses (QGEM) to decoherence using 3 superpositions[J]. Phys. Rev. A 3, 032411 (2022).
\bibitem{31} Pan Li, Yi Ling, Zhangping Yu. Generation rate of quantum gravity induced entanglement with multiple massive particles, Phys. Rev. D 107 6, 064054 (2023).
\bibitem{26} W D{\"u}r, Vidal G, Cirac J I. Three qubits can be entangled in two inequivalent ways[J]. American Physical Society, 2000(6).
\bibitem{16} Wei T C, Goldbart P M. Geometric measure of entanglement and applications to bipartite and multipartite quantum states[J]. Physical Review A 68(4):4343-4349 (2003).
\bibitem{17} Zhu H, Chen L, Hayashi M. Additivity and non-additivity of multipartite entanglement measures[J]. New Journal of Physics, 2010, 12(8):2099-2154.
\bibitem{8} Yusef Maleki and Alireza Maleki, Complementarity-Entanglement Tradeoff in Quantum Gravity, arXiv:220501967v1 (2022).
\bibitem{12} A. F. Radkowski. Some aspects of the source description of gravitation. Annals of Physics 56, 319 (1970).
\bibitem{14} Bjerrum-Bohr N, Donoghue J. F, Holstein B R. Quantum gravitational corrections to the nonrelativistic scattering potential of two masses[J]. Physical Review D 67(8):875-875 (2003).
\bibitem{6} Chevalier H, Paige A J, Kim M S. Witnessing the non-classical nature of gravity in the presence of unknown interactions, arXiv: 2005.13922v1 (2020).
\bibitem{7} Guff T, N Boulle, Pikovski I. Optimal Fidelity Witnesses for Gravitational Entanglement, arXiv: 2112.08564v1 (2021).
\bibitem{25} Bin Yi, Urbasi Sinha, Dipankar Home, Anupam Mazumdar, Sougato Bose. Spatial qubit entanglement witness for quantum natured gravity. arXiv: 2211.03661v1 (2022).
\bibitem{3} Hall M, Reginatto M. On two recent proposals for witnessing nonclassical gravity[J]. Journal of Physics A Mathematical and Theoretical, (2017).
\bibitem{4} M. Kemal D{\"o}ner, Andr\'{e} Gro$\beta$ardt. Is gravitational entanglement evidence for the quantization of spacetime? arXiv: 2205.00939v1 (2022).
\bibitem{27} Sab\'{i}n C, Garc\'{i}a-Alcaine G. A classification of entanglement in three-qubit systems. Eur. Phys. J. D48 435 (2008).
\end{thebibliography}

\end{document}